\documentclass[useAMS]{mn2e}
\usepackage{graphicx}
\usepackage[authoryear]{natbib}

\newcommand{\um}{$\mu$m}

\newcommand{\spitzer}{{\it Spitzer}}
\newcommand{\filtsONIR}{{$u^{*}Bg'Ri'Iz'ZJHK$}}

\newcommand{\sex}{\textsc{SExtractor}}
\newcommand{\hz}{\textsc{HyperZ}}
\newcommand{\scamp}{\textsc{Scamp}}
\newcommand{\swarp}{\textsc{Swarp}}

\title[The UV Luminosity Function at $0.6<z<1.2$]{The ultraviolet luminosity function of star-forming galaxies between redshifts of 0.6 and 1.2}

\author[M.J. Page et al.]
{M.J. Page$^{1}$, 
T. Dwelly$^{2}$,
I. McHardy$^{3}$,
N. Seymour$^{4}$,
K.O. Mason$^{5}$,
M. Sharma$^{1}$,
\and
J.A. Kennea$^{6}$,
T.P. Sasseen$^{7}$,
J.I. Rawlings$^{1}$,
A.A. Breeveld$^{1}$, 
I. Ferreras$^{1,8,9}$,
\and
N.S. Loaring$^{1}$,
D.J. Walton$^{10}$,
M. Symeonidis$^{1}$
\\ 
\\
$^{1}$Mullard Space Science Laboratory, University College London,
Holmbury St Mary, Dorking, Surrey, RH5 6NT, UK\\
$^{2}$tdastro.com, Seymour Rd., Bath, BA1 6DY, UK\\
$^{3}$Department of Physics and Astronomy, University Southampton, 
Southampton SO17 1BJ, UK\\
$^{4}$International Centre for Radio Astronomy Research, Curtin University, Bentley WA 6102, Australia\\
$^{5}$Satellite Applications Catapult, Fermi Avenue, Harwell Campus, Didcot, Oxfordshire OX11 0QR, UK\\
$^{6}$Department of Astronomy and Astrophysics, The Pennsylvania State University, 525 Davey Laboratory, University Park, PA 16802, USA\\
$^{7}$AT\&T, 5383 Hollister Avenue, Santa Barbara, CA 93111, USA\\
$^{8}$Instituto de Astrof\'isica de Canarias, Calle V\'ia L\'actea s/n, E-38205, La Laguna, Tenerife, Spain\\
$^{9}$Departamento de Astrof\'\i sica, Universidad de La Laguna, E38206 La Laguna, Tenerife, Spain\\
$^{10}$Institute of Astronomy, University of Cambridge, Madingley Road, Cambridge CB3 0HA, UK\\
}

\begin{document}

\date{Accepted ----. Received ----; in original form ----}

\pagerange{\pageref{firstpage}--\pageref{lastpage}} 
\pubyear{2013}
\maketitle

\label{firstpage}

\begin{abstract}
  We use ultraviolet imaging taken with the {\em XMM-Newton} Optical
  Monitor telescope (XMM-OM), 
 covering 280~arcmin$^{2}$
  in the UVW1 band
  ($\lambda_{eff}=2910$\,\AA) to measure rest-frame ultraviolet
  1500\,\AA\ luminosity functions of galaxies with redshifts $z$
  between 0.6 and 1.2. The XMM-OM data are supplemented by a large
  body of optical and infrared imaging to provide photometric
  redshifts. The XMM-OM data have a significantly narrower point-spread-function (resulting in less source confusion) and simpler K-correction
  than the {\em GALEX} data previously employed in this
  redshift range. Ultraviolet-bright active galactic nuclei are excluded to ensure that the luminosity functions relate directly to the star-forming galaxy population. Binned luminosity functions and parametric
  Schechter-function fits are derived in two redshift intervals:
  $0.6<z<0.8$ and $0.8<z<1.2$. We find that the luminosity function evolves 
  such that the characteristic absolute magnitude $M^{*}$ is brighter for $0.8<z<1.2$ than for $0.6<z<0.8$.
\end{abstract}

\begin{keywords}
  galaxies: evolution -- galaxies: luminosity function -- ultraviolet: galaxies
\end{keywords}

\section{Introduction}
\label{sec:introduction}

The luminosity function of galaxies is one of the most fundamental
measurements of the population. Ultraviolet light derives
predominantly from young stars, hence the ultraviolet luminosity
function (UVLF) relates directly to the 
distribution of unobscured star formation
in galaxies. 
The UVLF is well described by a Schechter function 
\citep{schechter76}, akin to galaxy luminosity functions at optical and 
near-infrared wavelengths \citep{sullivan00}. 

The Earth's atmosphere is opaque at wavelengths shorter than 3000~\AA, hence observations from space are required to probe this spectral region. In the nearby Universe, the UVLF has been derived primarily from surveys
carried out with NASA's {\em GALEX} satellite \citep{martin05}. 
The far-ultraviolet (FUV) channel of {\em GALEX}, in particular,
provides photometry in a broad passband centred at approximately
1500~\AA; this wavelength range is ideally placed for measuring the
emission from young, massive stars which have lifetimes $<$~100~Myr,
which in turn are a direct tracer of star-formation \citep{kennicutt12}. 
{\em GALEX} has
surveyed large areas of the sky in both its FUV channel and its longer
wavelength near-ultraviolet (NUV) channel.  In combination with
redshift surveys, these data have been used to produce luminosity
functions of low redshift ($z<0.6$) galaxies that extend several
magnitudes fainter than the characteristic absolute magnitude of the luminosity function $M^{*}$
\citep{wyder05,arnouts05}. 

At $z>1.2$, rest-frame 1500~\AA\ falls in
the optical to near-IR spectral range in the observer frame, and is
accessible with ground-based as well as space-based
instruments. Again, luminosity functions which extend several
magnitudes fainter than $M^{*}$ have been produced for the redshift
range $1.2<z<4$ \citep[e.g. ][]{reddy09, parsa16}.

Between $z=0.6$ and $z=1.2$, studies of the UVLF are somewhat more
difficult.  In this redshift range, {\it GALEX}'s passbands fall to
the blue of rest-frame 1500~\AA, and {\it GALEX} becomes hampered by
source confusion, such that it can not be used to probe much fainter
than $M^{*}$. Furthermore, these redshifts are not high enough to
place the 1500~\AA\ UV into the optical window, so ground-based
facilities can not be used to measure directly the rest-frame
1500~\AA\ emission. 

An important distinction should be made
between direct measurements of the UVLF, in which the galaxies that
are counted are found in images with wavelengths corresponding
approximately to rest-frame 1500~\AA, and indirect measurements of the
UVLF, in which the galaxies that are counted are found in images that
correspond to wavelengths longer than rest-frame 1500~\AA, and their
luminosity function is constructed by extrapolation of their
magnitudes to shorter wavelengths.  A half-way house between these two
approaches is represented by studies in which the galaxies to be
counted are found in images that correspond to wavelengths longer than
rest-frame 1500~\AA, but for which the photometry used to calculate their
absolute magnitudes is obtained from images
corresponding approximately to 1500~\AA\ in the rest-frame.

Beginning with the direct measurements,
\citet{arnouts05} provide some measurements based on
{\em GALEX}, in the redshift ranges 0.6-0.8 and
0.8-1.2. More recently, \citet{oesch10} used the UV channel of the Wide
Field Camera 3 on the {\em Hubble Space Telescope} to push the UVLF to
fainter absolute magnitudes, reaching $M_{1500}=-17.0$ in the redshift
range $0.5<z<1.0$. Despite these works, constraints on the faint end
slope $\alpha$ and characteristic magnitude $M^{*}$, which define the
shape of the Schechter function, remain quite crude for redshifts
between 0.6 and 1.2. Indeed, the somewhat surprising situation prevails 
that there are better {\em direct} measurements of the UVLF at $z>6$ 
\citep[e.g.][]{bouwens15,ishigaki18}, the epoch of reionization, 
than there are in the relatively recent cosmic past ($0.6<z<1.2$).

The studies by \citet{cucciati12} and \citet{moutard20} derived
indirect measurements of the UVLF covering the redshift interval
$0.6<z<1.2$, where the rest-frame 1500~\AA\ absolute magnitudes are
extrapolated from longer wavelength measurements. Compared to the
direct measurement of the UVLF in this redshift range, these
ground-based studies benefit from much larger statistical samples, but
the accuracy of their measurements depends critically on the fidelity
of the photometric extrapolation into the UV, and therefore on the
fitting software and library of spectral templates that is used.

Sitting somewhere between these two approaches, lies the study of
\citet{hagen15}, who constructed luminosity functions using a deep
exposure of the {\em Chandra} Deep Field South with the Ultraviolet
and Optical Telescope (UVOT) on the {\em Neil Gehrels Swift
  Observatory}. Their galaxy sample is selected in the UVOT U-band,
and hence at longer wavelengths than rest-frame 1500~\AA\ for $z<1$,
but with UVOT photometry at shorter wavelengths permitting precise
determination of the rest-frame 1500~\AA\ absolute
magnitudes. Finally, it should be noted that part of the study of
\citet{moutard20} also falls into this half-way house category of
measurements. \citet{moutard20} used GALEX measurements for the
brightest, $z<0.9$ galaxies in their sample, and hence the
corresponding parts of their UVLFs are based on direct measurements of
absolute magnitude.

\begin{figure}
\begin{center}
\includegraphics[width=55mm, angle=270]{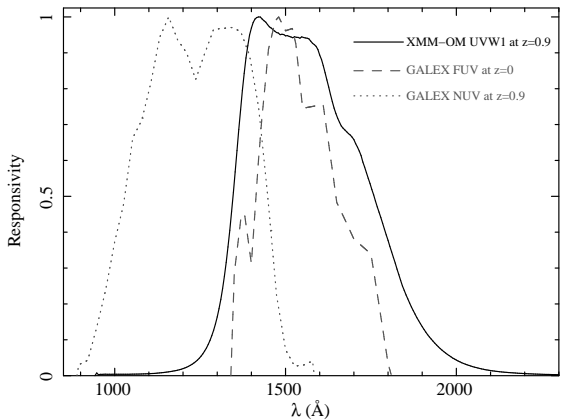}
\caption{The rest-frame responsivity of the XMM-OM UVW1 passband at $z=0.9$ 
compared to that of the {\em GALEX} FUV passband at $z=0$ 
and the {\em GALEX} NUV passband at $z=0.9$.
}
\label{fig:UVW1_FUV}
\end{center}
\end{figure}

In this paper we use an observation of the 13$^{H}$ {\em XMM-Newton}
Deep Field \citep{mchardy03,loaring05} taken with the {\em XMM-Newton}
Optical Monitor \citep[XMM-OM; ][]{mason01} through the UVW1 filter, 
which has an effective wavelength of 2910\AA,
to examine the UV luminosity function of galaxies in the redshift
interval $0.6<z<1.2$. The 13$^{H}$ Field is centred at
13$^{h}$~34$^{m}$~30$^{s}$~$+37^{\circ}$~53\arcmin\ (J2000), and
corresponds to an area of exceptionally low Galactic extinction
\citep[E(B-V)=0.005 mag;][]{schlafly11}.  This low extinction, and the
availability of redshifts facilitated by extensive multiwavelength
follow-up, make the 13$^{H}$ Field an excellent location for a study
of the UV galaxy luminosity function.  The XMM-OM UVW1 passband is
ideal to probe the rest-frame 1500~\AA\ emission in this redshift
range: at $z=0.9$ it covers a range of rest-frame wavelengths 
similar to the {\em GALEX} FUV passband at $z=0$, 
whereas for $z\ge 0.9$ the {\em GALEX} NUV passband probes 
shorter rest-frame wavelengths (Fig.~\ref{fig:UVW1_FUV}).
The XMM-OM has a much smaller point spread
function than {\em GALEX}: the full width, half maximum of the XMM-OM
with the UVW1 filter is just 2.0~arcsec \citep{ebrero19} compared to
5.3~arcsec for the {\em GALEX} NUV channel \citep{morrissey07}. In
this regard, XMM-OM also has an advantage over the Swift UVOT, which
has a full width, half maximum of 2.4~arcsec for its UVW1 filter 
\citep{breeveld10}.  To
our knowledge, this work is the first use of the XMM-OM to measure the
UVLF of galaxies.

As part of this paper, we describe the optical and infrared imaging of
the 13$^{H}$~Field and the techniques that were used to derive
photometric redshifts. This material serves also as a reference for
the data and techniques used in earlier works on the 13$^{H}$~Field
that make use of these photometric redshifts
\citep{seymour09,symeonidis09,seymour10}.

The paper is laid out as follows.  In Section \ref{sec:observations}
we describe the ultraviolet XMM-OM imaging which forms the basis of
this study, and the supporting optical and infrared data which were
used to derive redshifts.  The methods used to construct the
luminosity function are described in
Section~\ref{sec:construction}. Our results are presented in
Section~\ref{sec:results}. Section~\ref{sec:discussion} contains our
discussion, and our conclusions are presented in
Section~\ref{sec:conclusions}. Appendix~A describes some analysis of
the supporting optical and infrared image properties that was required
for the photometric redshift determination.

Throughout this paper magnitudes are given in the AB system 
\citep{oke83}. 
We have assumed cosmological parameters
$H_{0}=70$~km~s$^{-1}$~Mpc$^{-1}$, $\Omega_{\Lambda}=0.7$ and
$\Omega_{\rm m}=0.3$. Unless stated otherwise, uncertainties are
given at 1\,$\sigma$.

\section{Observations and data reduction}
\label{sec:observations}

Our UV luminosity functions are based on a catalogue of sources
detected in an XMM-OM UVW1 image, together with spectroscopic and
photometric redshifts for the sources. Therefore the UVW1 imaging, 
which has an effective wavelength of 2910\AA,
is supported by optical spectroscopic observations and a coherent suite
of optical, near-infrared and mid-infrared imaging, from which
high-quality photometric redshifts can be derived. The observations
are described below; a summary of the imaging is given in
Table~\ref{tab:phot_obs_table}.

\subsection{XMM-OM ultraviolet imaging}
\label{sec:xmmom}

The 13$^{H}$ field was observed with {\em XMM-Newton} over the course
of 3 orbits in June 2001 \citep{loaring05}. The XMM-OM UVW1
observations comprised four exposures in full-frame, low-resolution
mode of duration 5~ks each, for a total exposure of 20~ks.

\begin{figure}
\begin{center}
\includegraphics[width=80mm, angle=0]{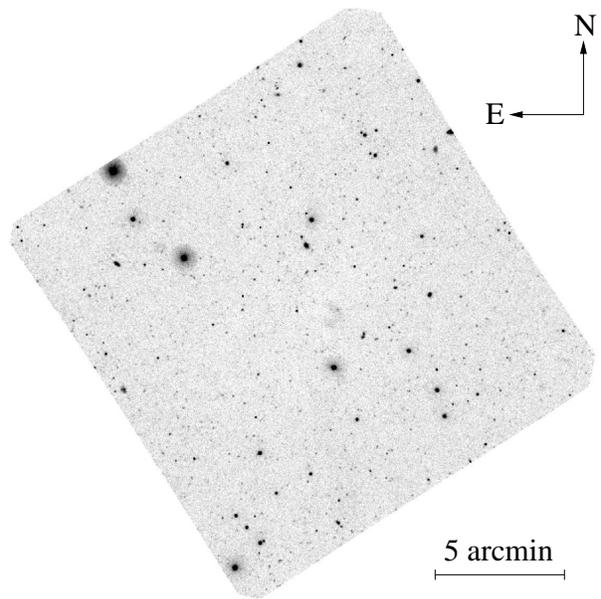}
\caption{Co-added 20~ks XMM-OM UVW1 image of the 13$^{H}$ field.}
\label{fig:13h_image}
\end{center}
\end{figure}

Initial reduction of the XMM-OM data was carried out using the
standard {\em XMM-Newton} Science Analysis Software (SAS) version 12.0 
task {\sc omichain}, as far as the correction of each of the individual 5~ks
exposures for modulo-8 noise via the SAS task {\sc ommodmap}. Then
each exposure was corrected for background scattered light structure
by subtracting a background template derived from many different
XMM-OM fields observed through the UVW1 filter, and replacing it with a
uniform background level at the mean of the subtracted template. The
images were then corrected for distortion and aligned with the
equatorial coordinate frame using the SAS task {\sc omatt}. The images
were registered to the Sloan Digital Sky Survey (SDSS) astrometric
reference frame by cross correlating source positions derived from the
XMM-OM images to the corresponding positions in the 
SDSS Data Release 6 \citep{adelmanmccarthy08}; the rms scatter between 
rectified XMM-OM and SDSS positions is 0.5~arcsec. Then, the
images were co-added using the SAS task {\sc ommosaic}. The resulting
image is shown in Fig.~\ref{fig:13h_image}, and covers a sky area of 280.1 
arcmin$^{2}$. 

Source detection and
photometry were carried out with the SAS task {\sc
  omdetect}. This task uses a sequence of peak-finding and
thresholding to find both point-like and extended sources. For
point-like sources, photometry is measured in an aperture of
5.7~arcsec radius for bright sources, or 2.8~arcsec radius for faint
sources, although intermediate aperture sizes are sometimes employed
for measuring close pairs of objects. 
For extended sources the photometric aperture consists of all
clustered pixels $>2 \sigma$ above the background level. For more
details of the source detection procedure, see \citet{page12}.
A total of 734 sources were detected in the UVW1 image at a 
signal-to-noise threshold of 3, with the faintest objects detected 
having UVW1 magnitudes of 24.3.

\subsection{Supporting optical spectroscopic observations}
\label{sec:spectra}

Optical spectroscopic observations provide precise redshifts. The
13$^{H}$ field has been used for extragalactic survey work for almost
three decades, and so has a long history of spectroscopic observations
targetting populations of sources selected at a variety of
wavebands. Table~\ref{tab:spectroscopy} provides basic information on 
the spectroscopic
campaigns that have furnished the majority of the redshifts used in
this study. Principally, the redshifts come from observations with
the William Herschel Telescope on La Palma, using the Autofib2
fibre-positioner together with the WYFFOS fibre-fed spectrograph, with
Gemini GMOS and Keck LRIS and DEIMOS observations extending the follow-up 
to the
faintest optical magnitudes. 

\begin{figure}
\begin{center}
\hspace{-8mm}
\includegraphics[angle=270,width=90mm]{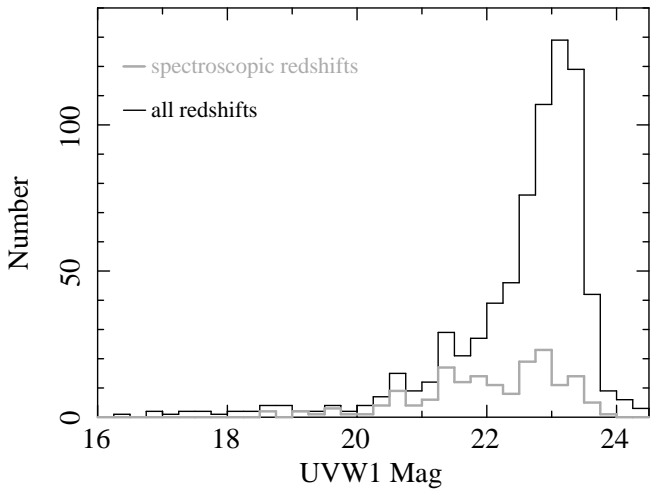}
\end{center}
\caption{UVW1 magnitude distribution of extragalactic UVW1 sources. 
The bold grey histogram 
corresponds to the sources with spectroscopic redshifts, while the black 
histogram corresponds to sources with spectroscopic or photometric redshifts.
}
\label{fig:magdistr}
\end{figure}

A total of 168 UVW1 sources in the 13$^{H}$ 
field have spectroscopic redshifts. Their UVW1 magnitude distribution 
is shown in Fig.~\ref{fig:magdistr}.

\begin{table*}
\caption{Summary of optical spectroscopic observations. }
\label{tab:spectroscopy}
\begin{tabular}{@{}llcc@{}cl}
\hline
Observatory & Instrument & Wavelength & Resolution & Dates          &  Notes \\
&&range (\AA)&(\AA)&&\\
\hline	
%&&&&&\\
WHT&AF2/WYFFOS&3800--9200&20&1998/04/30-1998/05/04&R300B grating, large fibres\\
Keck I&LRIS&3800--9250&6.9&2002/04/12--2002/04/14&400/8500 red grating + 600/4000 blue grism\\
WHT&AF2/WYFFOS&3800--9200&8.8&2002/05/09-2002/05/10&R300B grating, small fibres\\
Keck II&DEIMOS&4000--9500&3.5&2003/03/30--2003/03/31&600ZD grating\\
WHT&AF2/WYFFOS&3800--9200&8.1&2003/03/30-2003/04/01&R316R grating, small fibres\\
WHT&AF2/WYFFOS&3800--9200&8.8&2006/04/25-2006/04/26&R300B grating, small fibres\\
Gemini-N&GMOS&4050--9600&11.1&2007/01/12-2008/05/08&R150 grating, nod and shuffle\\
WHT&AF2/WYFFOS&3800--9200&8.8&2014/06/02-2014/06/03&R300B grating, small fibres\\
WHT&AF2/WYFFOS&4800--9200&8.1&2015/05/08-2015/05/10&R316R grating, small fibres\\
\hline
\end{tabular}
\end{table*}

\subsection{Supporting optical and infrared imaging}
\label{sec:optical_obs}

\begin{table*}
\caption{Summary of the optical, near-infrared, and mid-infrared imaging 
observations in the 13$^{H}$
field. $T_{tot}$ is the total exposure time (in seconds) spent on sky, and $T_{used}$ is the total exposure time of frames used in the the final stacks.}
\label{tab:phot_obs_table}
\begin{tabular}{@{}llllrrl}
\hline
Observatory & Instrument & Band & \multicolumn{1}{c}{Dates}          & $T_{tot}$    & $T_{used}$    &  Notes \\
\hline	
{\em XMM-Newton}      & XMM-OM & $UVW1$      & 2001/06/12--2001/06/24         & 20000         & 20000          &  \\ 
\hline
Subaru      & SuprimeCam & $R$      & 2000/12/24--2000/12/25         & 5400         & 2400          &  central pointing, Chips w67c1,w93c2 faulty\\ 
            &            & $R$      & 2003/05/02--2003/05/05         & 18450        & 17550         &  3x3 mosaic\\ 
            &            & $B$      & 2004/04/17--2004/12/16         & 10800        & 7200          &  Only Dec 2004 data useful\\ 
            &            & $I$      & 2004/12/11                     & 3000         & 3000          &  \\ 
            &            & $z'$     & 2004/04/17                     & 4550         & 3150          &  \\ 
\hline
CFHT        & MegaCam    & $u^{*}$  & 2004/05/11--2005/04/06         & 20786        & 20786         & Single pointing\\  
            &            & $g'$     & 2004/05/12--2005/07/10         & 21606        & 21606         & \\  
            &            & $i'$     & 2004/05/09--2004/07/21         & 10752        & 10752         & \\  
\hline
INT         & WFC        & $Z$      & 2006/03/03--2006/03/09         & 115200       & 115200        & 2x2 mosaic\\  
\hline
CFHT        & WIRCam     & $J$      & 2007/05/05--2007/07/08         & 17360        & 17360         & 2x2 mosaic\\  
            &            & $H$      & 2006/04/09--2007/07/13         & 31110        & 31110         & 2 epochs of data\\  
\hline
UKIRT       & WFCAM      & $K$      & 2006/06/02--2006/06/06         & 45480        & 45480         & Filled tile \\  
\hline
\spitzer    & IRAC       & all      & 2005/06/15                     & 36525           & 36525            & \\
\hline
\end{tabular}
\end{table*}

Here we describe the optical to infrared imaging observations
that were used to 
derive photometric redshifts and to select targets for our optical 
spectroscopic observations.
A total of
14 bands from $u^*$ to 8.0\um\ are used here.  
The observational details are
summarised in Table~\ref{tab:phot_obs_table}.

\subsubsection{CFHT-Megacam $u^{*}$, $g'$, and $i'$ data}

We observed the 13$^{H}$ field using the CFHT-MEGACAM wide field
camera during 2004 and 2005. Totals of 5.8, 6.2 and 3.0 hours
of useful exposure time were obtained in the $u^{*}$, $g'$ and $i'$
bands respectively.  Fully calibrated stacked images and weight maps
were downloaded from the MegaPipe
website\footnote{http://www.cadc-ccda.hia-iha.nrc-cnrc.gc.ca/en/megapipe/}.
The MegaPipe reduced images are photometrically and astrometrically
tied to the SDSS imaging of the field.

\subsubsection{Subaru SuprimeCam $B$, $R$, $I$, and $z'$ data}
\label{sec:subaru}
Observations of the 13$^{H}$ field were made using
Subaru-SuprimeCam \citep{miyazaki02}.
The first epoch of
SuprimeCam imaging was carried out in the $R$-band in December 2000 
\citep{mchardy03}, and 
further $R$ band imaging was obtained
in 2003. 
$B$, $I$, and $z'$
band observations were obtained between April and December 2004.
For each epoch of imaging several jittered images were obtained per band to
fill in the gaps between the individual 
CCD chips and to aid cosmic ray rejection.

Our reduction strategy drew heavily on the techniques 
described 
in detail by \citet{erben05} and \citet{gawiser06}. We
used a combination of standard IRAF tools
to debias, flatfield and (for $I$ and $z'$) remove the fringing. We then used
TERAPIX \citep{bertin02} and our own tools to
calibrate the data astrometrically and photometrically, and to make
the final stacked images.

\subsubsection{INT-WFC $Z$ band data}
We observed the 13$^{H}$ field in the $Z$ band using the INT-WFC
instrument over seven nights in March 2006.  
The WFC data were detrended (debiased, flatfielded, superflattened) using
standard IRAF tools. Special attention was required to mitigate the
effects of the strong and variable fringing (5--10\%\ of the sky
level) seen in the $Z$ band. 
The final
image stack was made using \scamp\ and \swarp\ tools, combining a total
of around 25 hours of good data. The astrometry and photometry of the
$Z$ band image were tied to the $z'$ measurements of stars and galaxies in the SDSS-DR6 catalogue.

\subsubsection{CFHT-WIRCam $J$ and $H$ band data}
We obtained observations of the 13$^{H}$ field in service mode with
CFHT-WIRCam in the J and H bands 
during the 2006A and 2007A semesters.
The data were preprocessed using CFHT's `iwii' WIRCam preprocessing
pipeline.
The TERAPIX team provided image stacks \citep{marmo07}. 
The zero-points of the WIRCam
images were tied to the 2MASS imaging in the 13$^{H}$ field.

\subsubsection{UKIRT-WFCAM $K$ band data}
We carried out imaging of the 13$^{H}$ field in the K band with UKIRT-WFCAM
during June 2006.  A total of 21\,hours were obtained over 5 nights in
good seeing conditions.  
Preprocessed and calibrated `interleaved' stacks and weight
images were obtained from the WSA
archive\footnote{http://surveys.roe.ac.uk/wsa/index.html}. These images were
combined to create a single stacked image and weight map using the
\scamp\ and \swarp\ tools. 
The photometric calibration of the final stack was derived from
the zeropoints of the calibrated WSA data, which are derived from
on-sky measurements of standard stars interspersed between the science
observations.

\subsubsection{\spitzer\ IRAC 3.6--8.0\um\ data}
A $\sim$30$\times$60\,arcmin stripe covering the 13$^{H}$ field was observed 
with \spitzer\ \citep{werner04} during June and July 2005.  
Data were obtained in all four IRAC 
bands \citep[3.6\um, 4.5\um, 5.8\um\ and 8.0\um;][]{fazio04}. The exposure per
pixel is approximately 500s in each band. 
The IRAC basic calibrated data were processed using the standard {\em Spitzer} {\sc mopex} package \citep{makovoz05} to generate a mosaiced science image for each IRAC band. The standard {\em Spitzer} photometric calibration was adopted.

\subsection{Determination of optical and infrared image characteristics}

Several aspects of the images were characterised before we obtained
the multi-band photometry that was used to derive photometric
redshifts. The point spread function (PSF) of each image was measured
using bright, but not saturated, stars in the image, and aperture corrections derived. The
limiting magnitude for each image, as a function of aperture size, was
determined by analysis of the noise properties within randomly placed 
circular apertures. The bandpasses of the images were derived from the
available information on the optical properties of the telescopes,
instruments, and atmospheric extinction. Then, the zeropoints were
fine-tuned by fitting stellar templates to the spectral energy
distributions of Galactic stars in the images. Each of these steps is
described in more detail in Appendix~\ref{app:characteristics}.

\subsection{Multiband photometry method}
We created a pipeline to make multi-band photometric measurements of all
objects detected in the optical and infrared imaging in the 13$^{H}$ field. 
First, a master catalogue of optical and infrared detections was created using 
 \sex\ \citep{bertin96} to construct separate
catalogues from the images in each filter. 
\sex\ was configured to record
MAG\_AUTO, MAGERR\_AUTO, FLUX\_RADIUS and FLAGS parameters for each
source (FLUX\_RADIUS records the radius which contains 50\% of the
source flux). 

The individual \sex\ catalogues were then merged band by band into a
master catalogue containing one row per {\em unique} source.  The
master catalogue is built up filter by filter and source by source.
Detections across multiple bands are considered to be the same source
if they lie within a small cross-matching radius (0.8~arcsec
for the majority of the optical and NIR images, 1.0~arcsec for J-band,
1.2~arcsec for IRAC 3.6$\mu$m and 4.5$\mu$m, and 1.4 arcsec for IRAC
5.8$\mu$m and 8$\mu$m). The catalogue merging was ordered such that the 
deepest and most complete wavebands were added first, and the shallowest, 
least complete wavebands were added last. The position of any source 
detected across
multiple bands was `refined' by taking the signal-to-noise-weighted
average of the individual positions of the source in each optical/NIR
filter where it is significantly detected. These position refinements
are typically small ($<0.1$~arcsec) but ensure that the position
determined in any single band does not disproportionately affect the
combined source position.

Aperture photometry in each band was then carried out at the location
of each source in the master catalogue. The procedure utilised \sex\
in double image mode, where the `detection' image is a synthetic image
made with point sources placed at the locations of each master
catalogue object. Aperture corrections were applied to account for the
different PSFs in different passbands; see 
Appendix~\ref{app:characteristics} for more details. 
In order to maximize the signal
to noise ratio in the photometry of faint sources, and to reduce the
aperture correction uncertainties for brighter, resolved galaxies, we
have used different sized apertures depending on the apparent $R$
magnitude of each source.  For objects brighter than $R = 18$ we use a
5~arcsec diameter aperture, for objects with $18<R<20$ a
3~arcsec aperture, and for fainter objects we use a 2~arcsec
diameter aperture.

\subsection{Photometric Redshift Method}
\label{sec:photoz_method}

We have used the \hz\ photometric redshift fitting package
\citep{bolzonella00}.
We experimented with a number of sets of model galaxy SED templates,
including the default template sets provided with \hz\ (based on
GISSEL98 synthesis models), the \citet{coleman80} set, and the 
\citet{bruzual03} templates. 
Of
those we tried, we found that the galaxy template set presented in
\citet{rowan-robinson08} was best able to reproduce the spectroscopic
redshifts of galaxies in the 13$^{H}$ field.  The
\citet{rowan-robinson08} set 
consists of seven galaxy templates ({\em E, E1, Sab, Sbc, Scd, Sdm, sb}) 
plus three active galactic nucleus (AGN) templates.  
Extinction was modelled using
the \citet{calzetti00} reddening law, with $A_V$ gridded in steps of
0.1 ranging up to 5.0 for the {\em sb} template, up to 2.0 for the
{\em Sdm} template, up to 1.0 for the {\em Sbc, Scd} and the three AGN
templates, and no extinction allowed for the {\em E, E1} and {\em Sab}
galaxy templates.
The absolute $R$ band magnitude 
was permitted
to range over $-27 < M_R < -16$.  

For the purposes of running \hz\ we assumed zero Galactic redenning
because the image zeropoints have been calibrated against de-reddened
stars which we assume lie behind the $E(B-V)=0.006$ of Galactic dust
\citep{schlegel98}\footnote{https://irsa.ipac.caltech.edu/applications/DUST/}
seen in the direction of the 13$^{H}$ field.
The magnitude uncertainties were increased by 0.05 mags in the IRAC
bands to account for the zero-point and aperture correction
uncertainties of the IRAC data. A minimum magnitude error threshold of
0.01 was adopted for all measurements in all filters. Our treatment
for photometric measurements fainter than the nominal $1\sigma$
magnitude limit was to replace the measured flux with zero, and set
the flux error to the $1\sigma$ value for the band in question.

The SEDs of AGN and starbursts at restframe wavelengths longer than
5\um\ may become complicated by PAH and silicate features, as well as
continuum emission from hot dust heated by an AGN. Therefore we have excluded IRAC
5.8 and 8.0\um\ data from the photometric redshift fits for most objects.  However,
for faint, high redshift ($z\gg1$) galaxies, the longer wavelength IRAC
bands become useful as they can constrain the position of the
redshifted 1.6\um\ stellar bump.  Therefore we include IRAC 5.8 and
8.0\um\ data in the photometric redshift fit only for optically faint galaxies
($R>24$), that have useful detections (magnitude error in 5.8 or
8.0\um\ bands $<0.3$) and have flat or decreasing SEDs in the IRAC
bands. Specifically we require that $[5.8] + 0.3 > [4.5]$ OR $[8.0] +
0.3 > [5.8]$.

\subsection{Photometric redshift fidelity}
\label{sec:photoz_fidelity}

\begin{figure}
\begin{center}
\hspace{-8mm}
\includegraphics[angle=0,width=90mm]{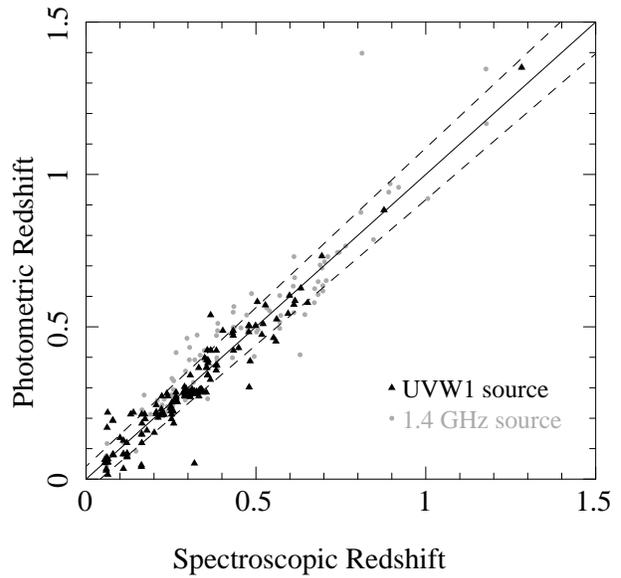}
\end{center}
\caption{Photometric redshift against spectroscopic redshift for galaxies in the 13$^{H}$ field that have spectroscopic redshifts. 146 UVW1-selected galaxies (from this work) are shown as black triangles, and 174 1.4~GHz radio-selected galaxies \citep{seymour08} are shown as grey dots. The solid line shows the one-to-one relation (solid) and the dashed lines indicate $\delta z = \pm 0.042$ (see Section~\ref{sec:photoz_fidelity}). 
}
\label{fig:specz_vs_photoz}
\end{figure}

A comparison of our photometric redshifts to the
redshifts of spectroscopically identified galaxies in the
13$^{H}$ field is shown in Fig. \ref{fig:specz_vs_photoz}. Broad-line AGN 
(quasi-stellar objects and Seyfert 1 galaxies) are not shown, because continuum 
variability compromises photometric redshifts for such objects in studies such as ours, when 
the imaging in the different bands is not contemporaneous.
For the 146 UVW1-detected galaxies which have spectroscopic redshifts,  the
photometric redshift residuals ($\delta z = [z_{photo} - z_{spec}]/[1+z_{spec}]$, where $z_{photo}$ is the photometric redshift and $z_{spec}$ is the spectroscopic redshift)
have a RMS $\sigma_{\delta z} = 0.042$ and a mean,
$\overline{\delta z} = -0.005\pm0.003$.  
This scatter is comparable to the photo-$z$ accuracy achieved in other deep photometric
redshift studies
\citep[e.g.][]{babbedge04,rowan-robinson08,ilbert06,mobasher07}. Adopting the definition from \citet{ilbert06} for a catastrophic failure of the photometric redshift as $|\delta z| > 0.15$, we find only one catastrophic failure out of the 146 UVW1-detected galaxies with spectroscopic redshifts.

The distribution of UVW1 magnitudes for sources with spectroscopic redshifts can be compared to the overall magnitude distribution of extragalactic sources in Fig.~\ref{fig:magdistr}. While the spectroscopic sources span almost the full range of UVW1 magnitudes, the median UVW1 magnitude for galaxies (excluding broad-line AGN) with spectroscopic redshifts is 22.2, compared to 23.0 for galaxies with only photometric redshifts. Given this difference in the median magnitudes, we have derived the RMS $\sigma_{\delta z}$ separately for three UVW1 magnitude intervals to see how the scatter in photometric redshift changes with magnitude. For the magnitude intervals 21$<$UVW1$\le$22, 22$<$UVW1$\le$23 and UVW1$>$23, we obtain $\sigma_{\delta z}$ of 0.046, 0.042 and 0.045 respectively. It appears that the accuracy of the photometric redshifts changes little with UVW1 magnitude to the limit of our UVW1-selected sample.

\subsection{Association of UVW1 sources 
with optical counterparts}
\label{sec:association}

To match the UVW1
sources to counterparts in the optical we have used our deep imaging
in Johnson $B$ taken with the SuprimeCam on the Subaru Telescope 
(see Section~\ref{sec:subaru}). 
UVW1 sources were matched to the brightest $B$-band source within
2~arcsec. This matching radius is similar to the PSF of the UVW1
images, and to the $3\,\sigma$ positional error of XMM-OM sources once
systematics related to the distortion correction are taken into
account, and so represents a good compromise between maximising the
completeness of the associations and minimising the number of
incorrect associations. The median offset between UVW1 positions and
those of their optical counterparts is 0.43~arcsec, and 95~per cent of
the offsets are smaller than 1.25~arcsec.

\subsection{Assignment of redshifts}
\label{sec:redshiftassignment}

UVW1 sources were attributed with the redshifts of the optical
counterparts assigned in Section~\ref{sec:association}. Where
available, spectroscopic redshifts were used in preference to
photometric redshifts.

The list of UVW1-selected galaxies used for the construction of luminosity functions, together with photometry and redshifts, is given in Table~\ref{tab:sourcelist}.

\begin{table}
\caption{UVW1-selected galaxies used to construct the luminosity functions. The positions given are those derived from the UVW1 image. UVW1 mag is the UVW1 apparent magnitude in the AB system. The column labelled $z$ gives the redshift for the source, and the column labelled spec/phot indicates whether the redshift is derived from spectroscopic or photometric data. Note that only the first five lines are included in the paper; the full table is available as supplementary material.}
\label{tab:sourcelist}
\begin{tabular}{@{}c@{\hspace{2mm}}ccc@{\hspace{2mm}}c@{}}
\hline
RA&dec&UVW1 mag&$z$&spec/phot\\
\multicolumn{2}{c}{(J2000)}&&&\\
%                                                                                 RA_1                 DEC_1              
\hline
&&&&\\
13 33 47.81 &  +37 53 08.7 &   $23.169\pm0.226$ &    0.986  &  phot   \\ %   203.44921217909246   37.885737779461536 
13 33 50.09 &  +37 52 39.2 &   $24.326\pm0.358$ &    0.738  &  phot   \\ %   203.4587171697382    37.87754200441795  
13 33 53.42 &  +37 54 40.7 &   $22.945\pm0.159$ &    0.602  &  phot   \\ %   203.47258312612271   37.91130582543706  
13 33 53.87 &  +37 53 18.9 &   $22.882\pm0.178$ &    0.855  &  phot   \\ %   203.4744589788509    37.88858359076976  
13 33 55.23 &  +37 52 49.0 &   $23.227\pm0.204$ &    1.084  &  phot   \\ %   203.48012695799386   37.880291062192725 
\end{tabular}
\end{table}

\subsection{Exclusion of broad-line AGN}
\label{sec:AGN}

Quasi Stellar Objects (QSOs) and Seyfert 1 galaxies are AGN characterised by broad emission lines and bright ultraviolet continua. Their ultraviolet radiation is powered by accretion onto their central supermassive black holes rather than originating in stars or stellar processes. The motivation to construct ultraviolet galaxy luminosity functions is to characterise the properties of star-forming galaxies, rather than AGN, and hence it is important to exclude these broad-emission-line AGN from the luminosity functions. In particular, because AGN can reach much higher ultraviolet luminosities than the stellar emission from galaxies, their inclusion would significantly distort the shape of the luminosity function at the brightest absolute magnitudes. Hence we exclude all objects spectroscopically identified as AGN with broad (full width half maximum $> 1000$~km~s$^{-1}$) emission lines from the source list used to construct luminosity functions.

Fortunately, we are able to exclude these broad-line AGN quite thoroughly in the 13$^{H}$ {\em XMM-Newton} Deep Field. Their broad emission lines render them easier to identify and obtain redshifts through optical spectroscopy than other galaxies of comparable optical magnitudes, particularly at redshifts larger than 0.8. The original purpose of the 13$^{H}$ field was an X-ray and radio survey, primarily to study AGN emitting in these bands. As a result, AGN candidates have been the highest priority targets over many years of our spectroscopic follow-up campaigns. 
Five broad-line AGN with spectroscopic redshifts between 0.6 and 1.2 were excluded from our sample through this process.

As a further check for AGN contamination of the sample, we searched
for UVW1 sources which are not spectroscopically identified, but which
are within 2 arcseconds of an X-ray source detected in our {\em
  Chandra} imaging observations \citep{mchardy03}. We find three such
sources. Their photometric redshifts are between 1 and 1.2, and their
implied 0.5--7~keV X-ray luminosities at these redshifts are larger
than $10^{43}$~erg~s$^{-1}$, higher than any known star-forming galaxy
and implying that all three are AGN. Their implied UV absolute
magnitudes range from -20.4 to -21.4, at the bright end of the
luminosity function where AGN contamination could have a significant
impact on luminosity function shape. In all three sources there is a
significant possibility that the UV emission is dominated by an AGN,
and hence we excluded them from the sample.

\section{Construction of the luminosity function} 
\label{sec:construction}

\subsection{Completeness}
\label{sec:completeness}

The completeness of the UVW1 source detection process as a function of
magnitude was assessed by repeatedly injecting fake sources into the
UVW1 image, repeating the source detection process, and recording the
fraction of the injected sources which are recovered. The fake sources
were given Gaussian spatial profiles with FWHM equivalent to the XMM-OM PSF, 
i.e. point-like sources. At
the magnitudes and redshifts of interest ($z>0.6$, UVW1 magnitude $>
21$), almost all sources appear point-like to the {\sc omdetect}
source-detection software, and the 2.8~arcsec minimum-radius aperture used
in {\sc omdetect} to measure photometry renders the precise shape of the 
input source on
sub-PSF scales unimportant. The positions of the fake sources were
randomised over the image, and a single fake source was injected for
each source detection pass. A source was considered to have been
successfully recovered if a source was detected within 2 arcseconds of
the input position. A thousand injection/recovery trials were
performed at each input magnitude tested to build up a statistically
robust measurement of the completeness.

\begin{figure}
\begin{center}
\includegraphics[width=55mm, angle=270]{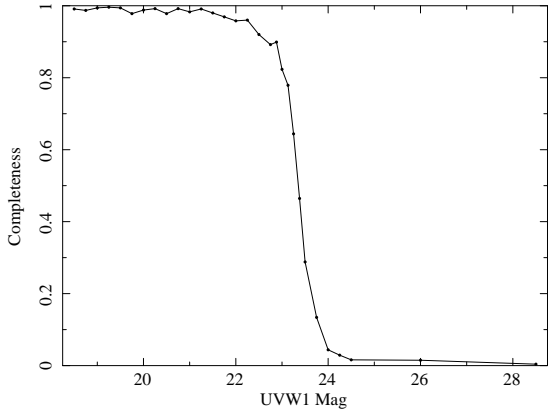}
\caption{Completeness of the source detection as a function of UVW1 magnitude, as determined from the simulations described in Section~\ref{sec:completeness}
}
\label{fig:completeness}
\end{center}
\end{figure}

The results are shown in
Fig.~\ref{fig:completeness}. The catalogue is 99 per cent complete for
UVW1 $\le 21$~mag, 85 per cent complete at UVW1~$=23$~mag, and falls
to below 7 per cent by UVW1~$=24$~mag. At the faint limit of the
trials, UVW1=28.5 mag, simulated sources contribute an inconsequential
number of counts to the image. The residual level of simulated source
recovery at this magnitude, 0.4 per cent, represents the
level of source confusion; at these magnitudes the
recovered sources are unrelated to the input sources, which are too
faint to be detected.

\subsection{Galactic extinction}
\label{sec:extinction}

The 13$^{H}$ field was chosen as an X-ray survey field because it has
an exceptionally low Galactic H\,I column density of around $6\times
10^{19}$~cm$^{-2}$ \citep{lockman86,branduardi94}. It therefore has a correspondingly low level of
Galactic extinction, which is beneficial for an extragalactic UV
survey field. To determine the reddening correction we have used the
extinction calibration from \citet{schlafly11} together with the dust
map of \citet{schlegel98}. The inferred Galactic reddening in UVW1 in
the direction of the 13$^{H}$ field is 0.027 mag, and all UVW1 magnitudes
have been corrected for this level of Galactic reddening.

\subsection{K-correction}
\label{sec:kcorrection}

\begin{figure}
\begin{center}
\includegraphics[width=55mm, angle=270]{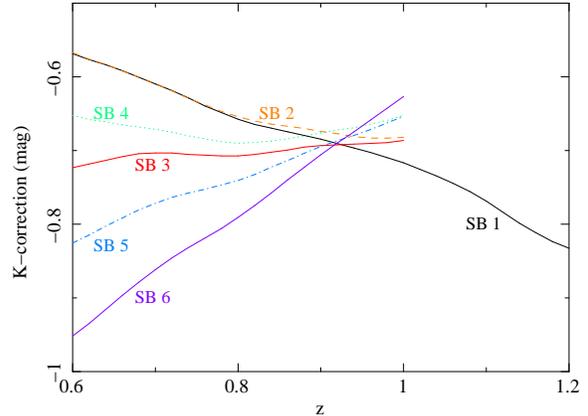}
\caption{K-corrections for the starburst templates from
\citet{kinney96} and \citet{calzetti94}, labelled as in
\citet{kinney96}. K-corrections for templates SB~2--6 end at $z=1$
because the templates do not extend below 1250~\AA. For template
SB~1, the spectrum of Mrk~66 from \citet{gonzalez98} has been used
to extend the template to shorter wavelengths, permitting K-corrections 
to be derived to $z=1.2$.  }
\label{fig:kcorr}
\end{center}
\end{figure}

K-correction, the correction of magnitudes from the observed
wavelength passband to a fixed rest-frame passband is a critical step
in the construction of luminosity functions, particularly in the
ultraviolet where extinction leads to a large variation in spectral
shape. For the reference rest-frame passband we have adopted the FUV
channel of {\em GALEX} which has a peak response close to 1500~\AA;
this choice ensures that our results can be directly compared to the
{\em GALEX}-derived UVLF of the local Universe
\citep{wyder05}. Fortunately, the choice of the UVW1 passband for our
XMM-OM observations (see Fig.~\ref{fig:UVW1_FUV}), and its proximity
to rest-frame 1500~\AA\ for the redshift range of interest in our study
($0.6<z<1.2$), leads to a very modest range of
K-correction. Fig.~\ref{fig:kcorr} shows K-corrections from the XMM-OM
UVW1 passband (in the observer frame) to the {\em GALEX}~FUV passband
(in the rest-frame) for the library of starburst templates presented
by \citet{kinney96} and \citet{calzetti94}. The template spectra have
a short-wavelength limit of 1250~\AA, and therefore cannot be used to
derive K-corrections beyond $z=1$. In order to facilitate K-correction
to larger redshifts, we have extended the \citet{kinney96} SB~1
template to shorter wavelengths using the spectrum of the
low-extinction starburst galaxy Mrk~66 from \citet{gonzalez98}. 
The K-corrections are almost
template-independent at $z=0.9$, and span a 0.4 mag range at the
low-redshift limit of our sample, $z=0.6$. 

\begin{figure}
\begin{center}
\includegraphics[width=55mm, angle=270]{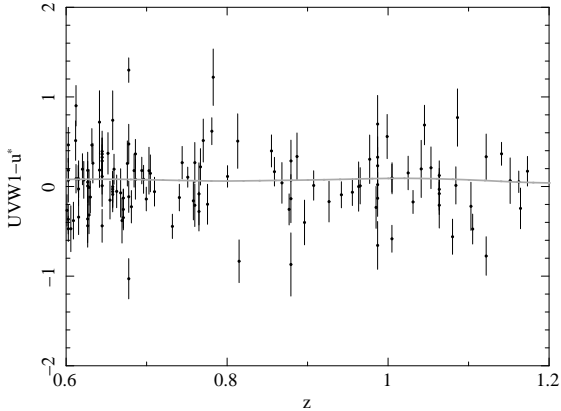}
\caption{UVW1$-u^{*}$ colours against redshift for the UVW1-selected galaxies in the 13$^{H}$ field (black data points) and the SB~1 galaxy template which is used for K-correction (grey curve).}
\label{fig:colours}
\end{center}
\end{figure}

UV selection favours low extinction galaxies, but to verify that the
SB~1 template is appropriate for the K-correction we have compared the
observed UVW1$-u^{*}$ colours of our galaxy sample with the
synthesized colours of the template over the redshift range of
interest. The results are shown in Fig.~\ref{fig:colours}. The
measurements are seen to be evenly distributed around the model curve
throughout the redshift range of interest. Quantitatively, the mean
UVW1$-u^{*}$ colours for the observed galaxies are
$\langle$UVW1$-u^{*}\rangle=0.06\pm0.04$ for $0.6<z<0.8$ and
$\langle$UVW1$-u^{*}\rangle=0.02\pm0.05$ for $0.8<z<1.2$, in excellent
agreement with the mean of the template curve, which is 0.07 in both
redshift ranges.

\subsection{Construction of  the binned luminosity functions}
\label{sec:binnedlf}

The method of \citet{page00} was used to construct binned
representations of the luminosity function. Two redshift ranges were
chosen, $0.6<z<0.8$ and $0.8<z<1.2$ to allow direct
comparison of our binned luminosity functions with those of
\citet{arnouts05} and \citet{hagen15}. 
For our survey, source completeness changes considerably between bright and faint magnitudes (Section~\ref{sec:completeness} and Fig.~\ref{fig:completeness}). Completeness below unity is equivalent to a proportional reduction in survey volume. To take this effect into account, the effective sky area used to compute the binned luminosity functions was obtained by multiplying the sky area of the UVW1 image (280.1 arcmin$^2$) by the source completeness in a series of discrete magnitude steps. The magnitude intervals and associated effective sky areas are given in Table~\ref{tab:skyarea}. 
Uncertainties on the binned 
luminosity functions were computed according to Poisson statistics using 
the approach described in \citet{gehrels86}.

\begin{table}
\caption{Effective sky area as a function of magnitude, used in the construction of the binned luminosity functions.}
\label{tab:skyarea}
\begin{tabular}{cc}
UVW1 magnitude range&Effective sky area \\
           (mag)&(arcmin$^{2}$)\\
\hline
&\\
$<18.50$     &279.0\\
$18.50-21.75$&276.8\\
$21.75-22.25$&274.3\\
$22.25-22.50$&271.8\\
$22.50-22.75$&260.3\\
$22.75-23.00$&243.0\\
$23.00-23.25$&185.0\\
$23.25-23.50$& 89.3\\
$23.50-23.75$& 43.2\\
\end{tabular}
\end{table}

\subsection{Measuring the Schechter function parameters}
\label{sec:fitting}

We used a parametric maximum likelihood fit to the data to estimate
the Schechter parameters $\alpha$ and $M^{*}$.  In the presence of
photometric errors on the magnitudes, the observed luminosity function
will be distorted from its original form in a manner analogous to the
distortion of source counts by measurement errors
\citep{eddington13}. The following approach is analogous to the
method developed by \citet{murdoch73} for radio source counts.

Suppose that for a source of true absolute magnitude $M$, 
the probability of obtaining a measured 
absolute magnitude in the interval $M'$ to
$M'+dM'$ is $P(M'|M,z)dM'$. It follows that the measured luminosity function
$\phi'(M')$ is related to the true luminosity function $\phi(M)$ by
\begin{equation}
\phi '(M') = \int P(M'|M,z)\phi(M)dM
\label{eq:phidash}
\end{equation}
The probability density $P_{i}$ of observing an object of measured 
absolute magnitude
$M'$ is 
\begin{equation}
P_{i}=\frac{\phi '(M_{i}',z_{i})}{\int \int \phi '(M',z) \frac{dV}{dz} dM' dz}
\label{eq:Pdash}
\end{equation}
The overall probability density of the observed distribution of objects is 
therefore 
\begin{equation}
P=\prod^{N}_{i=1} P_{i}
\end{equation}
where $N$ is the number of objects in the sample. This is equivalent to
minimising $C$ which is defined as
\begin{equation}
C= -2ln(P) = -2\sum^{N}_{i=1} ln(P_{i})
\label{eq:Cdash}
\end{equation}
Confidence regions can be
estimated by finding parameter values which give increased $C$, in the
same way that $\Delta \chi^{2}$ is used in $\chi^{2}$ fitting. The factor 2 in
equation \ref{eq:Cdash} 
is introduced so that the confidence intervals defined by 
$\Delta C$ are equivalent to those defined by $\Delta \chi^{2}$ 
\citep{lampton76}.

Substituting equation \ref{eq:Pdash} into equation \ref{eq:Cdash} and 
rearranging, we
obtain:
\[
C=2N\,ln \left(\int \int \int P(M'|M,z)\phi(M,z)dM \frac{dV}{dz} dM' dz\right)
\]
\begin{equation}
\ \ \ \ \ \ \ \ \ \ -2\sum^{N}_{i=1} ln\, 
\int P(M_{i}'|M,z_{i})\phi(M_{i},z_{i})dM
\label{eq:finalestimator}
\end{equation}

The probability distribution $P(M'|M,z)$ is equivalent to the probability distribution of observed apparent UVW1 magnitudes $m'$ around the true apparent UVW1 magnitude $m$ that corresponds to absolute magnitude $M$ at redshift $z$. The distribution of observed vs true apparent magnitudes can be obtained directly from the simulations that were used to derive the completeness in Section~\ref{sec:completeness}. For our implementation of Eq.~\ref{eq:finalestimator}, we constructed histograms of $m'-m$ at a series of fixed values of $m$, spaced by 0.25 mag. The histograms were linearly interpolated to obtain a distribution appropriate for arbitrary $m$. 
The table of effective sky area for specific magnitude ranges (Table~\ref{tab:skyarea}) was not employed for the maximum likelihood fitting. 
Instead, completeness in the source detection is taken into account naturally in the fitting, because the histograms of $m'-m$ are normalised by the number of input sources in the simulations, but contain only those sources which were detected. This is equivalent to setting $\int P(M'|M,z)dM$ equal to the completeness at the apparent magnitude $m$ corresponding to $(M,z)$. Volume calculations assumed the full sky area of the survey (280.1 arcmin$^{2}$) to the limiting UVW1 magnitude of the survey (24.3 mag).

In the maximum likelihood scheme that we have outlined, $C$ is not sensitive to the normalisation of the Schechter function $\phi^{*}$. Hence $\phi^{*}$ is not a fitted parameter, and is instead chosen such that the predicted number of objects in the sample is equal to the observed number, i.e. $\phi^{*}$ is chosen to satisfy 
\begin{equation}
\int \int \phi '(M',z) \frac{dV}{dz} dM' dz = N
\label{eq:normalisation}
\end{equation}
The value of $\phi^{*}$ is highly covariant with the parameters $\alpha$ and $M^{*}$. 
In addition, the normalisation of the luminosity function can vary significantly between pencil-beam surveys due to cosmic variance. Estimates for the fluctuations in the numbers of galaxies in our survey due to cosmic variance were obtained using the tools provided by \citet{trenti08} and \citet{moster11}.  
Uncertainties on $\phi^{*}$ are given as the sum in quadrature of the Poisson uncertainty on the sample size $N$, the uncertainty due to cosmic variance from \citet{trenti08} and the 1~$\sigma$ covariance of $\phi^{*}$ with $M^{*}$.
  
\section{Results}
\label{sec:results}

\begin{figure}
\begin{center}
\hspace{-0.5cm}
\includegraphics[width=88mm, angle=0]{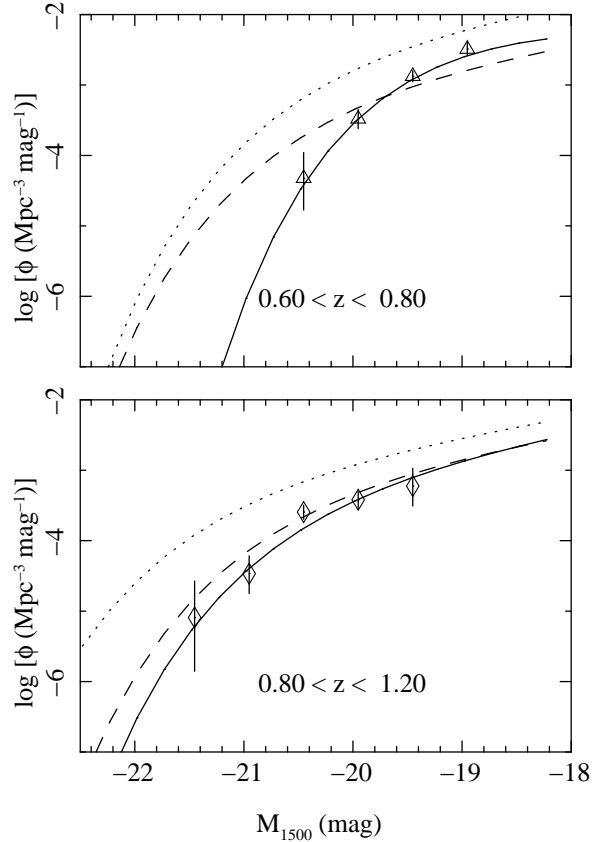}
\caption{UV luminosity function of galaxies in the redshift intervals $0.6<z<0.8$ (top panel) and $0.8<z<1.2$ (bottom panel). The datapoints show the binned luminosity functions derived from the 13$^{H}$ field as described in Section~\ref{sec:binnedlf}, and the solid curves show the best-fitting Schechter functions derived according to the method described in Section~\ref{sec:fitting}. For comparison, the dashed lines show the best fitting Schechter functions obtained by \citet{arnouts05}, and the dotted lines show the maximum-likelihood Schechter functions obtained by \citet{hagen15}.}
\label{fig:magi}
\end{center}
\end{figure}

Our binned luminosity functions, constructed as described in Section~\ref{sec:binnedlf}, are shown in Fig.~\ref{fig:magi}. The
bins are 0.5 mag wide in absolute magnitude ($M_{1500}$); the faintest
bins are centred at $M_{1500}=-18.95$ and $M_{1500}=-19.45$ for the
redshift ranges $0.6<z<0.8$ and $0.8<z<1.2$ respectively.  In the
$0.8<z<1.2$ redshift range our binned luminosity function appears
consistent with the Schechter function model obtained by
\citet{arnouts05}, but for $0.6<z<0.8$ our binned luminosity function
appears to be significantly steeper than the \citet{arnouts05} model. Tabulated values for the binned luminosity functions are provided in Table~\ref{tab:magi}.

\begin{table}
\caption{Binned luminosity function measurements. $M_{1500}$ is the centre of the absolute magnitude bin in the rest-frame {\em GALEX} FUV band; the absolute magnitude bins are 0.5 mag wide. $N$ is the number of galaxies in the bin. $\phi$ is the luminosity function.}
\label{tab:magi}
\begin{tabular}{cccc}
Redshift range&$M_{1500}$ & $N$ & Log~$\phi$ \\
           &(mag)&&(log [Mpc$^{-3}$~mag$^{-1}$])\\
\hline
$0.6<z<0.8$& $-20.45$ & 2 & $-4.32^{+0.37}_{-0.45}$\\
$''$       & $-19.95$ & 13 & $-3.48^{+0.13}_{-0.14}$\\
$''$       & $-19.45$ & 35 & $-2.88\pm 0.08$\\
$''$       & $-18.95$ & 25 & $-2.49^{+0.09}_{-0.10}$\\
&&&\\
$0.8<z<1.2$&  $-21.45$ & 1 & $-5.09^{+0.52}_{-0.76}$\\
$''$       & $-20.95$ & 4 & $-4.47^{+0.25}_{-0.28}$\\
$''$       & $-20.45$ & 23 & $-3.59\pm 0.10$\\
$''$       & $-19.95$ & 15 & $-3.42^{+0.12}_{-0.13}$\\
$''$       & $-19.45$ & 4 & $-3.22^{+0.25}_{-0.28}$\\
\end{tabular}
\end{table}

\begin{figure}
\begin{center}
\hspace{-0.5cm}
\includegraphics[width=85mm, angle=-90]{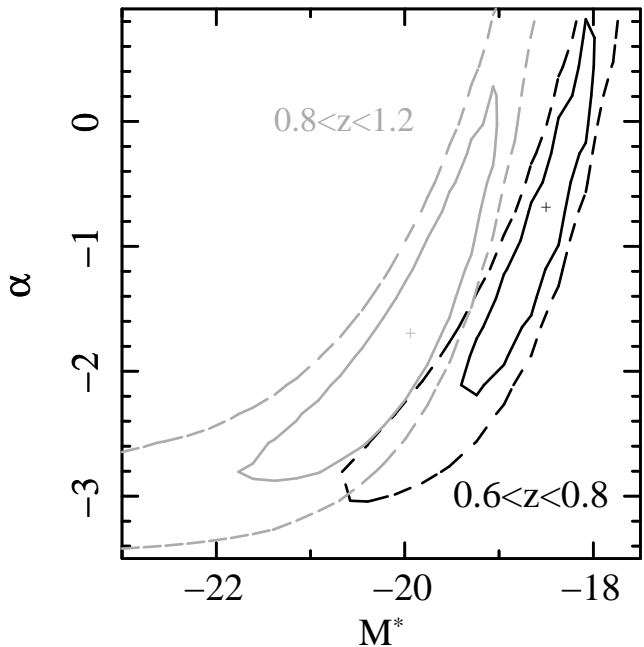}
\caption{Confidence contours for the fitted Schechter parameters $M^{*}$ and $\alpha$ in the redshift intervals $0.6<z<0.8$ (black) and $0.8<z<1.2$ (grey). The solid and dashed contours correspond respectively to 68 and 95 per cent confidence for two interesting parameters.}
\label{fig:contours}
\end{center}
\end{figure}

The results from our maximum likelihood model fitting, as described in Section~\ref{sec:fitting}, are recorded in
Table~\ref{tab:results}, and the confidence contours for $\alpha$ and
$M^{*}$ are shown in
Fig.~\ref{fig:contours}. As is inevitable
for a Schechter function fit to data which does not probe far into the
faint, power law section, 
the confidence contours are elongated, showing
significant covariance between $\alpha$ and $M^{*}$. 
According to the online tool\footnote{Calculations were carried out assuming completeness of 50 per cent and a halo filling factor of 0.5.} provided by \citet{trenti08} cosmic variance contributes to the uncertainty on $\phi^{*}$ at the level of 17 per cent in the redshift range $0.6<z<0.8$, and 15 per cent for $0.8<z<1.2$. The tool provided by \citet{moster11} gives slightly higher estimates, at 20 and 18 per cent respectively.

\begin{table}
\caption{Schechter function parameters from maximum likelihood fitting. $N$ is the number of galaxies included in the fit. 
$M_{lim}$ gives the absolute magnitude in the rest-frame {\em GALEX} FUV band that corresponds to the limiting apparent magnitude in our survey (UVW1=24.3) at the central redshift of the relevant redshift interval.
}
\label{tab:results}
\begin{tabular}{lccc@{\hspace{1mm}}c@{\hspace{1mm}}c}
Redshift & $N$ & $\alpha$ & $M^{*}$ & $\phi^{*}$ & $M_{lim}$\\
interval           &&&(mag)&($10^{-3}$~Mpc$^{-3}$)&(mag)\\
%           &&&&&\\
\hline
%&&&\\
$0.6-0.8$& 77  & $-0.7\pm1.1$ & $-18.5^{+0.4}_{-0.6}$ & $10.5^{+2.2}_{-5.5}$ & -18.27\\
$0.8-1.2$& 50  & $-1.7^{+1.2}_{-0.8}$&$-19.9^{+0.6}_{-0.9}$ & $1.2^{+0.9}_{-1.1}$& -19.11\\
\end{tabular}
\end{table}

\section{Discussion}
\label{sec:discussion}

We have constructed binned UV luminosity functions in the redshift
ranges $0.6<z<0.8$ and $0.8<z<1.2$, and carried out maximum-likelihood
fitting of Schechter function models to the unbinned data. To our
knowledge, ours is the first study since \citet{arnouts05} to provide
independent constraints on the UV luminosity function parameters in
these two redshift ranges using a UV-selected sample of galaxies.

Our luminosity functions can be compared directly with the best fit
models in the same redshift ranges derived by \citet{arnouts05} and
\citet{hagen15} in Fig.~\ref{fig:magi}. As \citet{arnouts05} and
\citet{hagen15} carried out their studies in different regions of the
sky to us, and to each other, differences in the normalisation, and to
a lesser extent $M^{*}$, between the three studies are expected
because of cosmic variance \citep{trenti08, moster11}. Ideally, cosmic
variance would be overcome with the use of a statistical sample of
independent UV survey fields, but at present there are only three. The
study of \citet{arnouts05} is based on a larger area of sky than our
study, or that of \citet{hagen15}, and so we might expect it to probe
the most representative range of large-scale-structure environment. A
simple estimate of the relative richness of our survey region compared
to that of \citet{arnouts05} can be obtained by comparing the models
around the faint limit of our survey, where we measure the largest
space density of galaxies. In the redshift range $0.6<z<0.8$, at
$M_{1500}=-19$, our model for log $\phi$ is higher than that of
\citet{arnouts05} by 0.2. If we were to assume that this difference
represents an overdensity in the 13$^{H}$ field due to cosmic
variance, comparison with Fig.~7 of \citet{trenti08} suggests that we
might expect our measurement of $M^{*}$ to be biased toward brighter
absolute magnitudes by about 0.2~mag. We consider a potential bias of
this size to be benign, because it is only half as large as the
1~$\sigma$ statistical uncertainty on $M^{*}$. In the redshift range
$0.8<z<1.2$, at $M_{1500}=-19.5$ our model for log $\phi$ differs from
that of \citet{arnouts05} by only 0.05, so we have no reason to expect
any significant bias in our determination of $M^{*}$.

\begin{figure}
\begin{center}
\includegraphics[width=80mm, angle=0]{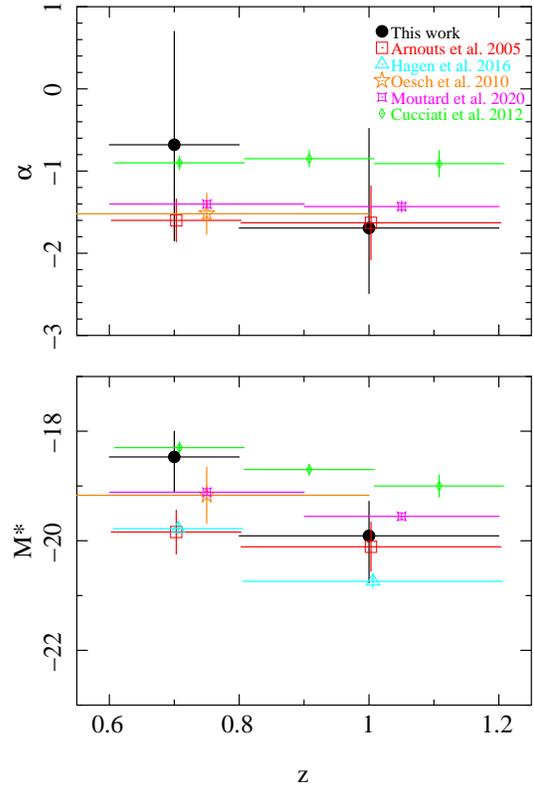}
\caption{Measurements of Schechter function parameters $\alpha$ and
$M^{*}$ from this work (closed black circles) and other UV surveys (open
symbols). Data points from other surveys have been slightly offset
in redshift to improve clarity.  }
\label{fig:alpha_mstar_z}
\end{center}
\end{figure}

The best fit $\alpha$ and $M^{*}$ parameters from our maximum
likelihood fitting, as well as those from other UV surveys covering
similar redshift ranges, are shown in
Fig.~\ref{fig:alpha_mstar_z}. The figure includes results from the
studies of \citet{cucciati12} and \citet{moutard20}, which we would
describe as indirect measurements in the sense that they are based on
surveys carried out at longer wavelengths from which UV luminosity
functions are obtained by extrapolation to shorter wavelengths, albeit
through sophisticated methods. These two works utilise large samples
of galaxies compared to the direct studies, and so their results have
relatively small statistical uncertainties. However, their
measurements of $\alpha$ and $M^{*}$ show a similar degree of
discrepancy with respect to each other as the measurements from the
direct studies; consequently, in statistical terms they are highly
discrepant with each other. It would seem likely that systematics
outweigh statistics as the dominant source of uncertainty in these
indirect measurements, demonstrating the value of direct UV surveys to
provide the ground-truth in this redshift range.

It is evident from
Fig.~\ref{fig:magi} that our binned luminosity functions in the two redshift
ranges are different, with the $0.6<z<0.8$ luminosity function
appearing steeper than that of the $0.8<z<1.2$ redshift range. In the
maximum likelihood fitting, the confidence contours for the two
redshift ranges cover different regions in $\alpha$ and $M^{*}$
parameter space (Fig.~\ref{fig:contours}). In contrast, 
evolution between the two redshift
ranges can not be deduced from the \citet{arnouts05} study, either
from inspection of their binned luminosity functions, or from the
parameters derived from their maximum likelihood fitting.

Inspection of the Schechter function parameters reported in
Table~\ref{tab:results} and their corresponding uncertainties confirms
the impression given by Figs~\ref{fig:magi} and \ref{fig:contours} 
that the evolution corresponds to a dimming of
$M^{*}$ between $0.8<z<1.2$ and $0.6<z<0.8$; evolution in $\alpha$ is
not suggested by the fits.  With no prior assumption about $\alpha$, the
evolution in $M^{*}$ is significant at 2$\sigma$; if we were to assume
that $\alpha=-1.5$ in both redshift ranges, the significance of the
change in $M^{*}$ would rise to 3$\sigma$.

Compared to the {\em GALEX}-based measurements in the same redshift
range of \citet{arnouts05}, our sample is smaller, covering a smaller
sky area and with a shallower limiting magnitude. On the other hand,
while \citet{arnouts05} suffers from systematics related to source
confusion, this is a much smaller issue for our survey: the superior
point spread function of XMM-OM leads to minimal source confusion ($<
1$~per cent). 

A second issue in which we consider our XMM-OM survey to have an
advantage over the {\em GALEX} study of \citet{arnouts05}
is in the shape of the UVW1 bandpass compared to {\em GALEX} NUV for
measuring rest-frame 1500~\AA\ photometry at $z>0.6$. In the construction of
luminosity functions, the bandpass determines the K-correction, and so
the suitability of the bandpass relates directly to the systematics
associated with K-correction. A UVW1-based survey has the advantage
that K-corrections for different spectral shapes converge at $z=0.9$,
and the range of K-corrections is small for the redshift range
$0.6<z<1.2$; see Fig~\ref{fig:kcorr}. 

We can gauge the level of
systematics inherent in our simple, single-template K-correction
scheme by repeating our Schechter-function fit using a different
template.  This test is easily performed in the $0.6<z<0.8$ range, for
which we can derive K-corrections for any of our template spectra (see
Fig.~\ref{fig:kcorr}).  We therefore repeated our Schechter-function
fit, using the K-correction from the SB~6 template, the most different
to the SB~1 template used to derive the results presented in
Section~\ref{sec:results}. The best-fitting $M^{*}$ changes to
$-18.6$, a difference of only 0.1 mag with respect to our original
results, while the best-fit faint-end slope changes to
$\alpha=-1.5$. These parameter changes are smaller than the
corresponding 1~$\sigma$ statistical uncertainties for our study.

\citet{arnouts05} do not quantify the effect such systematics may have
on their study; nor is it possible for us to quantify the effect of
K-correction-related systematics on their study from the information
that they provide. In qualitative terms, for {\em GALEX} NUV, the
redshift at which K-corrections for different templates converge is
$z=0.5$, so the K-corrections for different spectral shapes will
diverge progressively as redshift increases throughout the $0.6<z<1.2$
range. Because {\em GALEX} NUV is a wide passband, Lyman~$\alpha$
falls within its sensitive wavelength range throughout
$0.6<z<1.2$, and the wide variety of line profiles will induce 
additional scatter in the K-correction. 
Furthermore, the Lyman limit enters the {\em GALEX} NUV
passband at $z=0.85$, so K-corrections above this redshift depend on
the escape fraction of Lyman continuum photons, about which little is
known for galaxies in this redshift range, and which is likely to vary
significantly between galaxies \citep{izotov18}. Hence K-corrections
for the {\em GALEX} NUV band could be a non-trivial source of
systematics in the construction of luminosity functions in this 
redshift range.

Our luminosity functions reach similar absolute magnitude limits to
the luminosity functions constructed by \citet{hagen15} in the same
redshift ranges, despite their {\em Swift} UVOT data having a much
longer UVW1 exposure than our XMM-OM data. \citet{hagen15} based their
study, which had somewhat broader goals than ours, on a master sample
which was selected in the UVOT {\em u} filter, and as a result the
faint UV absolute magnitude limits were set by the onset of
colour-dependent incompleteness.  In their Schechter-function model
fits, \citet{hagen15} fixed the faint end slope $\alpha$ to the best
fit values obtained by \citet{arnouts05}; given the covariance between
$\alpha$ and $M^{*}$, their measurements of $M^{*}$ are therefore not
strictly independent from those of \citet{arnouts05}. Nonetheless,
their fits support the picture implied by our study that $M^{*}$
evolves such that it is brighter at $0.8<z<1.2$ than at
$0.6<z<0.8$. Visual inspection of the binned luminosity functions of
\citet{hagen15} also suggests that $M^{*}$ evolves in this fashion
between the two redshift ranges.

In the redshift
interval ($0.5<z<1$), that overlaps both of the redshift ranges that
we have studied, \citet{oesch10} used the {\em Hubble Space Telescope}
to measure the UV luminosity function to fainter magnitudes than
\citet{arnouts05}, \citet{hagen15}, or us. \citet{oesch10} obtained
$\alpha=-1.52\pm0.25$, which is consistent at 2$\sigma$ with our
measurements in both redshift ranges, and those of \citet{arnouts05}.

One issue that deserves some attention is the potential contamination
of UV galaxy luminosity functions by AGN. As discussed in
Section~\ref{sec:AGN}, we have explicitly excluded from our luminosity
functions QSOs and other AGN which we consider are likely to dominate
the rest-frame UV emission of their hosts, using a combination of
optical spectroscopy and X-ray indicators. Such AGN are small in
number compared to star-forming galaxies: a total of eight AGN were
excluded from our study, compared to more than 120 star-forming
galaxies. However, several of these AGN have very bright absolute
magnitudes, and, if included in the luminosity function, would
contribute in absolute magnitude bins where star-forming galaxies are
rare or absent from the sample. For our sample, inclusion of these AGN
would make a material difference to the bright end of our $0.8<z<1.2$
luminosity function. We find that with the AGN included, the best fit
$M^{*}$ would brighten by more than a magnitude compared to that
reported in Table~\ref{tab:results}. The best fit $\alpha$ would
steepen to $-2.5$ and the shape of the confidence contours in the
lower panel of Fig~\ref{fig:contours} would change to imply much
tighter constraints on $\alpha$. Thus we find that the exclusion of
UV-bright AGN is a critical step in constructing UV luminosity
functions of galaxies.

With this finding in mind, it is interesting to note that the studies
of \citet{arnouts05} and \citet{hagen15} make no mention of QSOs or
AGN at all. \citet{cucciati12} do not explicitly say whether any kind of
AGN are excluded from their luminosity functions. \citet{moutard20} do
discuss QSOs as a contaminant in their luminosity functions, with the
added complication that their photometric redshifts are probably
wrong for QSOs. They do not describe the criteria by which they attempt to
exclude UV-bright AGN. \citet{oesch10} exclude point-like sources in
their {\em Hubble} imaging as a means of cleaning stars from their
sample; at least powerful QSOs may be excluded by this method, but
they do not discuss potential AGN contamination of their sample. In
future we consider that it would be helpful for authors to outline
explicitly the steps that they have taken to exclude UV-bright AGN
from the samples of galaxies that they use to study the UV galaxy
luminosity function, because this step has a significant impact on the
luminosity functions that result.

Having carried out this study, which we hope will serve as a pilot for
more extensive application of XMM-OM data to the measurement of galaxy
UV luminosity functions, we make note of the following.  Measurements
of the faint end slope $\alpha$ are rather poor between redshifts of
0.6 and 1.2 in all surveys. Averaging our measurement of $\alpha$ in
the redshift range $0.8<z<1.2$ with that of \citet{arnouts05}, we
still have a 1~$\sigma$ uncertainty of $\pm 0.4$, which is far from
ideal. Further measurements of the faint end slope are thus required
if we are to determine the manner in which this parameter evolves as
the Universe's peak epoch of star formation came to a close. Given
that the value of $\alpha$ is largely thought to be determined by
feedback processes resulting from star formation (principally from
supernovae and stellar winds), it could be argued that proper
observational constraints on $\alpha$ are essential if we are to test
models for the evolution of star formation with cosmic time. Better
measurements of $\alpha$ demand luminosity functions that reach
fainter absolute magnitude limits than the survey we have presented in
this paper. Therefore progress here requires significantly deeper
XMM-OM UVW1 observations than the 20~ks observation that formed the
primary data in our study. Deeper observations already exist which may
be suitable for this purpose, though the collection and analysis of 
the required ancillary datasets for redshifts is a major task. Furthermore, 
so long as {\em XMM-Newton}
continues operations there is potential for considerable improvements
in UVW1 exposure time of deep extragalactic survey fields.

From the results of our study, it would appear that $M^{*}$ evolves
significantly in the redshift interval $0.6<z<1.2$, but the precision
of our measurements should still be regarded as crude, particularly
taking into account the covariance between $M^{*}$ and
$\alpha$. Better statistics at bright absolute magnitudes are
essential for breaking down the statistical uncertainties on $M^{*}$,
and so larger sky area as well as deeper magnitude limits would be of benefit.  
Thanks to {\em XMM-Newton}'s long service, tens of square degrees of
extragalactic sky have already been observed with XMM-OM in UVW1
\citep{page12}, hence the {\em XMM-Newton} Science Archive could prove
a rich resource for such a purpose if it can be combined with suitable
redshift data.

\section{Conclusions}
\label{sec:conclusions}

We have used XMM-OM UVW1 imaging of the 13$^{H}$ extragalactic survey
field to study the rest-frame 1500~\AA\ luminosity function of
galaxies in the redshift ranges $0.6<z<0.8$ and $0.8<z<1.2$. This is,
to our knowledge, the first use of XMM-OM data to measure galaxy
luminosity functions. The XMM-OM data are supported by a large body of
optical and infrared imaging, as well as optical spectroscopy, to
provide redshifts for the UV sources.  Our binned luminosity functions
are noticeably different for the two redshift ranges, indicating that
the luminosity function evolved significantly during the corresponding
period of cosmic history.  We used maximum-likelihood fitting
to fit Schechter-function models to the data. Our fits indicate that the
characteristic break magnitude $M^{*}$ is brighter in the higher
redshift interval.  In contrast, evolution in this redshift
range could not be inferred from the {\em GALEX} luminosity functions
of \citet{arnouts05}, though the {\em Swift} UVOT-based study of
\citet{hagen15} did find some evidence that $M^{*}$ brightens with
redshift in this redshift range.  
We argue that a combination of deeper, and
wider-area, XMM-OM UVW1 imaging could form an excellent basis for
major improvement in our understanding of the UV galaxy luminosity
function between redshifts of 0.6 and 1.2.

\section*{Acknowledgments}
\label{sec:acknowledgments}

This work was supported by Science and Technology Facility Council
(STFC) grant numbers ST/N000811/1, ST/S000216/1. DJW acknowledges
support from an STFC Ernest Rutherford Fellowship.  Based on
observations obtained with {\em XMM-Newton}, an ESA science mission with
instruments and contributions directly funded by ESA Member States and
NASA.  The WIRCam observations were made through the OPTICON
program. This work is based in part on observations made with the
Spitzer Space Telescope, which is operated by the Jet Propulsion
Laboratory, California Institute of Technology under a contract with
NASA. The William Herschel Telescope and the Isaac Newton Telescope
are operated on the island of La Palma by the Isaac Newton Group of
Telescopes in the Spanish Observatorio del Roque de los Muchachos of
the Instituto de Astrof\'isica de Canarias.

\section{Data Availability}

The primary data underlying this article are available from the 
{\em XMM-Newton} 
Science archive at https://www.cosmos.esa.int/web/xmm-newton. Supplementary 
data underlying this article will be shared on reasonable request 
to the corresponding author.

\bibliographystyle{mn2e}

\begin{thebibliography}{}

\bibitem[Adelman-McCarthy et~al.(2008)]{adelmanmccarthy08}
Adelman-McCarthy J.K., et~al.,
2008, ApJS, 175, 297

\bibitem[Arnouts et~al.(2005)]{arnouts05}
Arnouts S., et~al., 
2005, ApJ, 619, L43

\bibitem[{Babbedge et~al.(2004)
{Babbedge}, {Rowan-Robinson},
  {Gonzalez-Solares}, {Polletta}, {Berta}, {P{\'e}rez-Fournon}, {Oliver},
  {Salaman}, {Irwin} \& {Weatherley}}]{babbedge04}
Babbedge T.~S.~R., {et~al.}, 2004, MNRAS, 353, 654

\bibitem[{{Bertin} \& {Arnouts}(1996)}]{bertin96}
{Bertin} E., {Arnouts} S., 1996, A\&AS, 117, 393

\bibitem[Bertin et~al.(2002)]{bertin02} Bertin E., Mellier Y.,
 Radovich M., Missonnier G., Didelon P., Morin, B, 2002, Proceedings of Astronomical Data Analysis Software and Systems XI, 
 Astronomical Society of the Pacific 
 Converence Series 281, eds Bohlender D., Durand D., Handley
 T.H., 228

\bibitem[{{Bolzonella} {et~al.}(2000){Bolzonella}, {Miralles} \&
  {Pell{\'o}}}]{bolzonella00}
{Bolzonella} M., {Miralles} J.-M., {Pell{\'o}} R., 2000, A\&A, 363, 476

\bibitem[Bouwens et~al.(2015)]{bouwens15}
Bouwens R.J., et~al., 
2015, ApJ, 803, 34

\bibitem[Branduardi-Raymont et~al.(1994)]{branduardi94}
Branduardi-Raymont G., et~al.,
1994, MNRAS, 270, 947

\bibitem[Breeveld et~al.(2010)]{breeveld10}
Breeveld A. A., et~al.,
2010, MNRAS, 406, 1687

\bibitem[Bruzual \& Charlot(2003)]{bruzual03}
Bruzual G., Charlot S., 
2003, MNRAS, 344, 1000

\bibitem[{{Calzetti} {et~al.}(2000){Calzetti}, {Armus}, {Bohlin}, {Kinney},
  {Koornneef} \& {Storchi-Bergmann}}]{calzetti00}
{Calzetti} D., {Armus} L., {Bohlin} R.~C., {Kinney} A.~L., {Koornneef} J.,
  {Storchi-Bergmann} T., 2000, ApJ, 533, 682

\bibitem[Calzetti, Kinney \& Storchi-Bergmann(1994)]{calzetti94}
Calzitti D., Kinney A.L., Storchi-Bergmann T., 
1994, ApJ, 429, 582

\bibitem[Castelli \& Kurucz(2003)]{castelli03}
Castelli F. \& Kurucz R.L., 
2003, Proceedings of IAU Symposium 210, Eds N.E. Piskunov, W.W. Weis, 
D.F. Gray, Astronomical Society of the Pacific, A20. 

\bibitem[Coleman, Wu \& Weedman(1980)]{coleman80}
Coleman G.D., Wu C.-C, Weedman D. W.,
1980, ApJS, 43, 393

\bibitem[Cucciati et~al.(2012)]{cucciati12}
Cucciati O., et~al., 
2012, A\&A, 539, A31

\bibitem[Ebrero et~al.(2019)]{ebrero19}
Ebrero J., et~al.,
2019, ``{\em XMM-Newton} Users Handbook'', 
Issue 2.17, (ESA: {\em XMM-Newton} SOC)

\bibitem[Eddington(1913)]{eddington13}
Eddington A.S.,
1913, MNRAS, 73, 359

\bibitem[{{Erben} {et~al.}(2005){Erben}, {Schirmer}, {Dietrich}, {Cordes},
  {Haberzettl}, {Hetterscheidt}, {Hildebrandt}, {Schmithuesen}, {Schneider},
  {Simon}, {Deul}, {Hook}, {Kaiser}, {Radovich}, {Benoist}, {Nonino}, {Olsen},
  {Prandoni}, {Wichmann}, {Zaggia}, {Bomans}, {Dettmar} \&
  {Miralles}}]{erben05}
{Erben} T., {et~al.}, 2005, Astronomische Nachrichten, 326, 432

\bibitem[Fazio et~al.(2004)]{fazio04}
Fazio G.G., et~al., 
2004, ApJS, 154, 10

\bibitem[{{Gawiser} {et~al.}(2006){Gawiser}, {van Dokkum}, {Herrera}, {Maza},
  {Castander}, {Infante}, {Lira}, {Quadri}, {Toner}, {Treister}, {Urry},
  {Altmann}, {Assef}, {Christlein}, {Coppi}, {Dur{\'a}n}, {Franx}, {Galaz},
  {Huerta}, {Liu}, {L{\'o}pez}, {M{\'e}ndez}, {Moore}, {Rubio}, {Ruiz}, {Toft}
  \& {Yi}}]{gawiser06}
{Gawiser} E., {et~al.}, 2006, ApJS, 162, 1

\bibitem[Gehrels(1986)]{gehrels86}
Gehrels N., 1986, ApJ, 303, 336

\bibitem[Gonz\'alez et~al.(1998)]{gonzalez98}
Gonz\'alez R.M., Leitherer C., Heckman T., Lowenthal J.D., Ferguson H.C., 
Robert C., 
1998, ApJ, 495, 698

\bibitem[Hagen et~al.(2015)]{hagen15}
Hagen L.M.Z., Hoversten E.A., Gronwall C., Wolf C., Siegel M.H., Page M., 
Hagen A.,
2015, ApJ, 808, 178

\bibitem[{{Hewett} {et~al.}(2006){Hewett}, {Warren}, {Leggett} \&
  {Hodgkin}}]{hewett06}
{Hewett} P.~C., {Warren} S.~J., {Leggett} S.~K., {Hodgkin} S.~T., 2006, MNRAS,
  367, 454

\bibitem[{{Ilbert} {et~al.}(2006){Ilbert}, {Arnouts}, {McCracken},
  {Bolzonella}, {Bertin}, {Le F{\`e}vre}, {Mellier}, {Zamorani}, {Pell{\`o}},
  {Iovino}, {Tresse}, {Le Brun}, {Bottini}, {Garilli}, {Maccagni}, {Picat},
  {Scaramella}, {Scodeggio}, {Vettolani}, {Zanichelli}, {Adami}, {Bardelli},
  {Cappi}, {Charlot}, {Ciliegi}, {Contini}, {Cucciati}, {Foucaud}, {Franzetti},
  {Gavignaud}, {Guzzo}, {Marano}, {Marinoni}, {Mazure}, {Meneux}, {Merighi},
  {Paltani}, {Pollo}, {Pozzetti}, {Radovich}, {Zucca}, {Bondi}, {Bongiorno},
  {Busarello}, {de La Torre}, {Gregorini}, {Lamareille}, {Mathez}, {Merluzzi},
  {Ripepi}, {Rizzo} \& {Vergani}}]{ilbert06}
{Ilbert} O., {et~al.}, 2006, A\&A, 457, 841

\bibitem[{{Ivezi{\'c}} {et~al.}(2007){Ivezi{\'c}}, {Smith}, {Miknaitis}, {Lin},
  {Tucker}, {Lupton}, {Gunn}, {Knapp}, {Strauss}, {Sesar}, {Doi}, {Tanaka},
  {Fukugita}, {Holtzman}, {Kent}, {Yanny}, {Schlegel}, {Finkbeiner},
  {Padmanabhan}, {Rockosi}, {Juri{\'c}}, {Bond}, {Lee}, {Stoughton}, {Jester},
  {Harris}, {Harding}, {Morrison}, {Brinkmann}, {Schneider} \&
  {York}}]{ivezic07}
{Ivezi{\'c}} {\v Z}., {et~al.}, 2007, AJ, 134, 973

\bibitem[Ishigaki et~al.(2018)]{ishigaki18}
Ishigaki M., Kawamata R., Ouchi M., Oguri M., Shimasaku K., Yoshiaki O., 
2018, ApJ, 854, 73

\bibitem[Izotov et~al.(2018)]{izotov18}
Izotov Y.I.,  Worseck G.,  Schaerer D., Guseva N.G., Thuan T.X., Fricke, V.A., 
Orlito/'a I.,
2018, MNRAS, 478, 4851

\bibitem[Kennicutt \& Evans(2012)]{kennicutt12}
Kennicutt R.C. \& Evans N.J., 
2012, ARA\&A, 50, 531

\bibitem[Kinney et~al.(1996)]{kinney96}
Kinney A.L., Calzetti D., Bohlin R.C., McQuade K., Storchi-Bergmann T., 
Schmitt H.R., 
1996, ApJ, 467, 38

\bibitem[Lampton, Margon \& Bowyer(1976)]{lampton76} 
Lampton M., Margon B., Bowyer S., 
1976, ApJ, 208, 177

\bibitem[Loaring et~al.(2005)]{loaring05}
Loaring N.S., et~al., 
2005, MNRAS, 362, 1371

\bibitem[Lockman, Jahoda \& McCammon(1986)]{lockman86}
Lockman F.J., Jahoda K., McCammon D., 
1986, ApJ, 302, 432

\bibitem[Makovoz \& Khan(2005)]{makovoz05}
Makovoz D., Khan L., 
2005, Proceedings of Astronomical Data Analysis Software and Systems XIV, eds P. Shopwell, M. Britton, R. Ebert, Astronomical Society of the Pacific Conference Series 347, 81

\bibitem[Marmo(2007)]{marmo07}
Marmo C., 2007, Proceedings of Astronomical Data Analysis Software and 
Systems XVI, 
Astronomical Society of the Pacific 
Converence Series 376, eds Shaw R.A., Hill F., Bell
D.J., 285

\bibitem[Martin et al.(2005)]{martin05}
Martin D.C., et al., 
2005, ApJ, 619, L1

\bibitem[Mason et~al.(2001)]{mason01}
Mason K. O., et~al.,
2001, A\&A, 365, L36

\bibitem[McHardy et~al.(2003)]{mchardy03}
McHardy I.M., et~al., 
2003, MNRAS, 342, 802

\bibitem[{{Miyazaki} {et~al.}(2002){Miyazaki}, {Komiyama}, {Sekiguchi},
  {Okamura}, {Doi}, {Furusawa}, {Hamabe}, {Imi}, {Kimura}, {Nakata}, {Okada},
  {Ouchi}, {Shimasaku}, {Yagi} \& {Yasuda}}]{miyazaki02}
{Miyazaki} S., {et~al.}, 2002, PASJ, 54, 833

\bibitem[{{Mobasher} {et~al.}(2007){Mobasher}, {Capak}, {Scoville}, {Dahlen},
  {Salvato}, {Aussel}, {Thompson}, {Feldmann}, {Tasca}, {Lefevre}, {Lilly},
  {Carollo}, {Kartaltepe}, {McCracken}, {Mould}, {Renzini}, {Sanders},
  {Shopbell}, {Taniguchi}, {Ajiki}, {Shioya}, {Contini}, {Giavalisco},
  {Ilbert}, {Iovino}, {Le Brun}, {Mainieri}, {Mignoli} \&
  {Scodeggio}}]{mobasher07}
{Mobasher} B., {et~al.}, 2007, ApJS, 172, 117

\bibitem[Morrissey et~al.(2007)]{morrissey07}
Morrissey P., et~al., 
2007, ApJS, 173, 682

\bibitem[Moster et~al.(2011)]{moster11}
Moster B.P., Somerville R.S., Newman J.A., Rix H.-W.,
2011, ApJ, 731, 113

\bibitem[Moutard et~al.(2020)]{moutard20}
Moutard T., Sawicki M., Arnouts S., Golob A., Coupon J., Ilbert O., 
Yang X., Gwyn S.,
2020, MNRAS, 494, 1894

\bibitem[Murdoch, Crawford \& Jauncey(1973)]{murdoch73} 
Murdoch H.S., Crawford D.F., Jauncey D.L., 
1973, ApJ, 183, 1

\bibitem[Oesch et.~al.(2010)]{oesch10}
Oesch P.A., et~al., 
2010, ApJ, 725, L150

\bibitem[{{Oke} \& {Gunn}(1983)}]{oke83}
{Oke} J.~B., {Gunn} J.~E., 1983, ApJ, 266, 713

\bibitem[Page \& Carrera(2000)]{page00}
Page M.J., Carrera F.J., 
2000, MNRAS, 311, 433

\bibitem[Page et~al.(2012)]{page12}
Page M.J., et~al.,
2012, MNRAS, 426, 903

\bibitem[Parsa et al.(2016)]{parsa16}
Parsa S., Dunlop J.S., McLure R.J., Mortlock A.,
2016, MNRAS, 456, 3194

\bibitem[Reddy \& Steidel(2009)]{reddy09}
Reddy N.A., Steidel C.C.,
2009, ApJ, 692, 778

\bibitem[{{Rowan-Robinson} {et~al.}(2008){Rowan-Robinson}, {Babbedge},
  {Oliver}, {Trichas}, {Berta}, {Lonsdale}, {Smith}, {Shupe}, {Surace},
  {Arnouts}, {Ilbert}, {Le F{\'e}vre}, {Afonso-Luis}, {Perez-Fournon},
  {Hatziminaoglou}, {Polletta}, {Farrah} \& {Vaccari}}]{rowan-robinson08}
{Rowan-Robinson} M., {et~al.}, 2008, MNRAS, 386, 697

\bibitem[Schlafly \& Finkbeiner(2011)]{schlafly11}
Schlafly E.F., Finkbeiner D.P., 
2011, ApJ, 737, 103

\bibitem[Schechter(1976)]{schechter76}
Schechter P., 
1976, ApJ, 203, 297

\bibitem[Schlegel, Finkbeiner \& Davis(1998)]{schlegel98}
Schlegel D.J., Finkbeiner D.P., Davis M., 
1998, ApJ, 500, 525

\bibitem[Seymour et~al.(2008)]{seymour08}
Seymour N., et~al., 
2008, MNRAS, 386, 1695

\bibitem[Seymour et~al.(2009)]{seymour09}
Seymour N., Huynh M., Dwelly T., Symeonidis M., Hopkins A., McHardy I.M.,
Page M.J., Rieke G., 
2009, MNRAS, 398, 1573

\bibitem[Seymour et~al.(2010)]{seymour10}
Seymour N., Symeonidis M., Page M.J., Huynh M., Dwelly T., 
McHardy I.M., Rieke G., 
2010, MNRAS, 402, 2666

\bibitem[Sullivan et~al.(2000)]{sullivan00}
Sullivan M., Treyer M.A., Ellis R.S., Bridges T.J., Milliard B., Donas J.,
2000, MNRAS, 312, 442

\bibitem[Symeonidis et~al.(2009)]{symeonidis09}
Symeonidis M., Page M.J., Seymour N., Dwelly T., Coppin K., McHardy I., 
Rieke G.H., Huynh M.,  
2009, MNRAS, 397,1728

\bibitem[Talavera(2011)]{talavera11}
Talavera A., 2011, 
Technical Report XMM-SOC-CAL-TN-0019 issue 6.0, {\it XMM-Newton} 
Optical and UV monitor (OM) Calibration Status, ESA;
http://xmm2.esac.esa.int/docs/documents/CAL-TN-0019.pdf

\bibitem[Trenti \& Stiavelli(2008)]{trenti08}
Trenti M., \& Stiavelli M., 
2008, ApJ, 676, 767

\bibitem[Werner et~al.(2004)]{werner04}
Werner M., et~al., 
2004, ApJS, 154, 1

\bibitem[{{Wolf} {et~al.}(2004){Wolf}, {Meisenheimer}, {Kleinheinrich},
  {Borch}, {Dye}, {Gray}, {Wisotzki}, {Bell}, {Rix}, {Cimatti}, {Hasinger} \&
  {Szokoly}}]{wolf04}
{Wolf} C., {et~al.}, 2004, A\&A, 421, 913

\bibitem[Wyder et~al.(2005)]{wyder05}
Wyder T.K., et~al.,
2005, ApJ, 619, L15

\end{thebibliography}

\appendix

\section{Analysis of optical and infrared images}
\label{app:characteristics}
In order to reach the photometric precision required for reliable
photometric redshifts, the point spread functions, limiting
magnitudes, and bandpasses of the optical and infrared images had to
be measured. These characteristics were then used together with
measurements of stars in the images to tweak the photometric
zeropoints. Each of these steps is described here.  A summary of
pertinent image characteristics is given in Table
\ref{tab:imagedetails2}.

\begin{table*}
\noindent{{\bf Table~\ref{tab:imagedetails2}} Summary characteristics of the 
UV,
optical and infrared images of the 13$^{H}$ field. $\lambda_{eff}$ is the effective wavelength and $\Delta \lambda$ is 
FWHM of the filter passband. AB offset gives the difference between Vega and AB magnitude systems for each band, in the sense AB~mag = Vega~mag + AB offset. $A$ is the sky area covered with high quality
imaging (defined as the sky area where the weight map is at least 50\%
of the median weight). IQ is the mean measured FWHM of point-like
sources in the image. $\langle \Delta \underline{r} \rangle$ is the
median position difference between objects detected in the 13$^{H}$ field and objects in
the SDSS-DR6 catalogue.  ApCorr is the correction factor required to
correct the flux measured in a 2~arcsec diameter aperture (3.8~arcsec for IRAC bands). 
The 3$\sigma$ Depth columns give the
faintest AB magnitude and flux density 
for a point source such that its total flux can be
measured with signal to noise ratio $\ge3$ using a 2\,arcsec diameter
aperture 
(3.8~arcsec for IRAC bands, 5.7-arcsec for UVW1). Note that for UVW1, these entries simply refer to the faintest magnitude detected in the image, while for the other bands they were determined following the procedure described in Section A2.
$\sigma_{ZP}$ is the estimated accuracy of the relative
photometric calibration in this band.  }
\label{tab:imagedetails2}
\begin{tabular}{@{}lrrrrrrrrrr}
\hline
Band & $\lambda_{eff}$  & $\Delta \lambda$ & AB offset &      $A$  &    IQ  & $\langle \Delta \underline{r} \rangle$  &    ApCorr & \multicolumn{2}{c}{3$\sigma$ Depth} & $\sigma_{ZP}$ \\
         &     \um\         &       \um\       &   mag     & deg$^{2}$ & arcsec &  arcsec                                 &           &  mag   & $\mu$Jy &     mag              \\
\hline
UVW1  &      0.291       &            0.079 &    1.362  &     0.078 & 2.61  & 0.43  &   $-$   & 24.3 &  0.69  & 0.05           \\
$u^{*}$  &      0.386       &            0.056 &    0.405  &     0.983 & 1.175  & 0.13  &   1.433   & 26.1 &  0.14  & 0.02              \\
$B$      &      0.437       &            0.074 &   -0.100  &     0.264 & 0.870  & 0.12  &   1.273   & 27.0 & 0.056  & 0.02              \\
$g'$     &      0.475       &            0.100 &   -0.098  &     0.969 & 1.018  & 0.12  &   1.345   & 26.6 & 0.082  & 0.01              \\
$R$      &      0.645       &            0.082 &    0.208  &     1.133 & 0.697  & 0.11  &   1.153   & 26.1 &  0.13  & 0.02              \\
$i'$     &      0.756       &            0.098 &    0.386  &     0.974 & 0.794  & 0.10  &   1.141   & 25.1 &  0.34  & 0.01              \\
$I$      &      0.788       &            0.096 &    0.439  &     0.263 & 0.683  & 0.10  &   1.128   & 25.8 &  0.18  & 0.02              \\
$z'$     &      0.906       &            0.094 &    0.530  &     0.260 & 0.824  & 0.10  &   1.234   & 24.9 &  0.42  & 0.02              \\
$Z$      &      0.906       &            0.103 &    0.534  &     1.063 & 1.302  & 0.13  &   1.778   & 23.7 &   1.2  & 0.05              \\
$J$      &      1.248       &            0.109 &    0.942  &     0.464 & 0.686  & 0.12  &   1.159   & 23.3 &   1.7  & 0.03              \\
$H$      &      1.616       &            0.200 &    1.372  &     0.473 & 0.670  & 0.14  &   1.187   & 22.9 &   2.5  & 0.03              \\
$K$      &      2.190       &            0.247 &    1.900  &     0.791 & 0.806  & 0.15  &   1.172   & 23.0 &   2.3  & 0.03              \\
3.6\um\  &      3.513       &            0.505 &    2.818  &     0.434 & 2.25   & 0.20  &   1.359   & 23.0 &   2.3  & 0.10              \\
4.5\um\  &      4.443       &            0.670 &    3.290  &     0.439 & 2.00   & 0.21  &   1.397   & 22.6 &   3.4  & 0.10              \\
5.8\um\  &      5.647       &            0.950 &    3.783  &     0.433 & 2.32   & 0.28  &   1.650   & 20.6 &   23   & 0.10              \\
8.0\um\  &      7.617       &            1.950 &    4.424  &     0.435 & 2.85   & 0.28  &   1.841   & 20.5 &   23   & 0.10              \\
\hline			 								   
\end{tabular}
\end{table*}

\subsection{Image quality and aperture corrections}
\label{sec:dataQ_apercorr}
Our ground based science images are each compiled by stacking a number
of exposures, each obtained under different seeing conditions. The
science images therefore have a complex point spread function (PSF)
which in general is unlikely to be well represented by a single
Gaussian.  However, a good knowledge of the PSF is necessary in order
to calculate aperture corrections, and to discriminate between
point-like and resolved objects. Therefore, in order to measure the
PSF for each filter, we have produced a model PSF for each filter by
stacking a large number of stellar objects detected in the science
images.  The following process was carried out for each optical/NIR
band.  Firstly, a suitable subset of stellar objects were selected
from the SDSS-DR6 catalogue.  These stars were restricted to a magnitude
range that is unsaturated in our imaging data.  Small 150x150 pixel
postage stamp images were cut out from around each star using \swarp.
Each postage stamp image was normalised to the total flux of the star
(measured from the science image).  These postage stamp images were
reprojected onto a common pixel frame, and then, using \swarp, were
median averaged to produce a single PSF image per band.
\sex\ is run on the resultant PSF postage stamp image to determine the
image FWHM as well as the fractional flux enclosed as a function of
aperture radius. This information is used to provide accurate aperture
corrections for the photometric measurements used in this work.
The FWHM of stellar sources in each band is listed in table
\ref{tab:imagedetails2}, as well as the multiplicative correction
factor for a 2~arcsec diameter aperture.  For the IRAC and MIPS
bands, a good calibration is available on the IPAC 
website\footnote{https://irsa.ipac.caltech.edu/data/SPITZER/docs/dataanalysistools/},
and so the default FWHM and aperture correction factors were adopted.

\subsection{Image sensitivity and limiting magnitudes}
In our final stacked images, the noise in adjacent pixels is somewhat
correlated, due to the inevitable resampling step during the stacking
process. Therefore the raw rms of pixel values, scaled up
to the aperture area gives an underestimate of the true fluctuations
due to sky noise. We have followed a similar technique to
\citet{gawiser06} to calculate the rms noise fluctuations in each
band. 
Essentially, \sex\ was used to measure the flux in a variety of different sized
apertures at random positions on the science image.
A histogram of the measured flux values was generated for
each aperture sizes, and a Gaussian was fitted to the
negative part of this histogram (the positive part contains
contributions from astronomical objects).  The widths of the Gaussians
were then plotted against aperture area, to derive the sigma--area relation.
The 3$\sigma$
limiting magnitudes for a 2~arcsec aperture for each image band are
listed in table \ref{tab:imagedetails2}.  Note that the faintest {\em
  detectable} source in each band will be somewhat fainter than these
estimates because the detection process requires a significant
enhancement over rather fewer pixels than covered by a 2~arcsec
aperture.  
 
\subsection{Determination of science filter bandpasses}
\label{sec:filter_bandpasses}
In order to derive photometric redshifts we must know the
effective transmission curves for each science filter.  The effective
bandpass of each filter is the product of not only the filter
transmission, but also the detector quantum efficiency (QE), the
reflectivity and transmission of the telescope plus instrument optics,
as well as the atmospheric extinction. Each of these components can, in
general, change the wavelength dependence of the effective transmission
curve.  For each science band we have downloaded the filter
transmission and detector QE information from the website of the relevant 
observatory, as well as the reflectivity/transmission of the telescope
and instrument optics where available. 
For each of the optical filters we have modified the transmission
curve to approximate the effects of atmospheric extinction.
To do this we take the SDSS atmospheric transmission
curve\footnote{http://www.sdss.org/dr5/instruments/imager/filters/},
suitable for 1 airmass at Apache Point Observatory (APO) and convert
it to the appropriate average airmass during each set of science
observations, taking into account the relative
altitudes of the different observatories.  For the NIR bands, we adopted the
atmospheric extinction curve supplied by UKIRT, produced using the
program IRTRANS4, and obtained from the UKIRT
website\footnote{https://www.ukirt.hawaii.edu/astronomy/utils/atmos-index.html}.

For each science filter between $u*$ and $H$ we combined each of these
components to generate an effective transmission curve.  For the WFCAM
K-band data the effective transmission curve of \citet{hewett06} was
adopted which already incorporates all the necessary information.

\begin{figure}
\begin{center}
\includegraphics[angle=0,width=80mm]{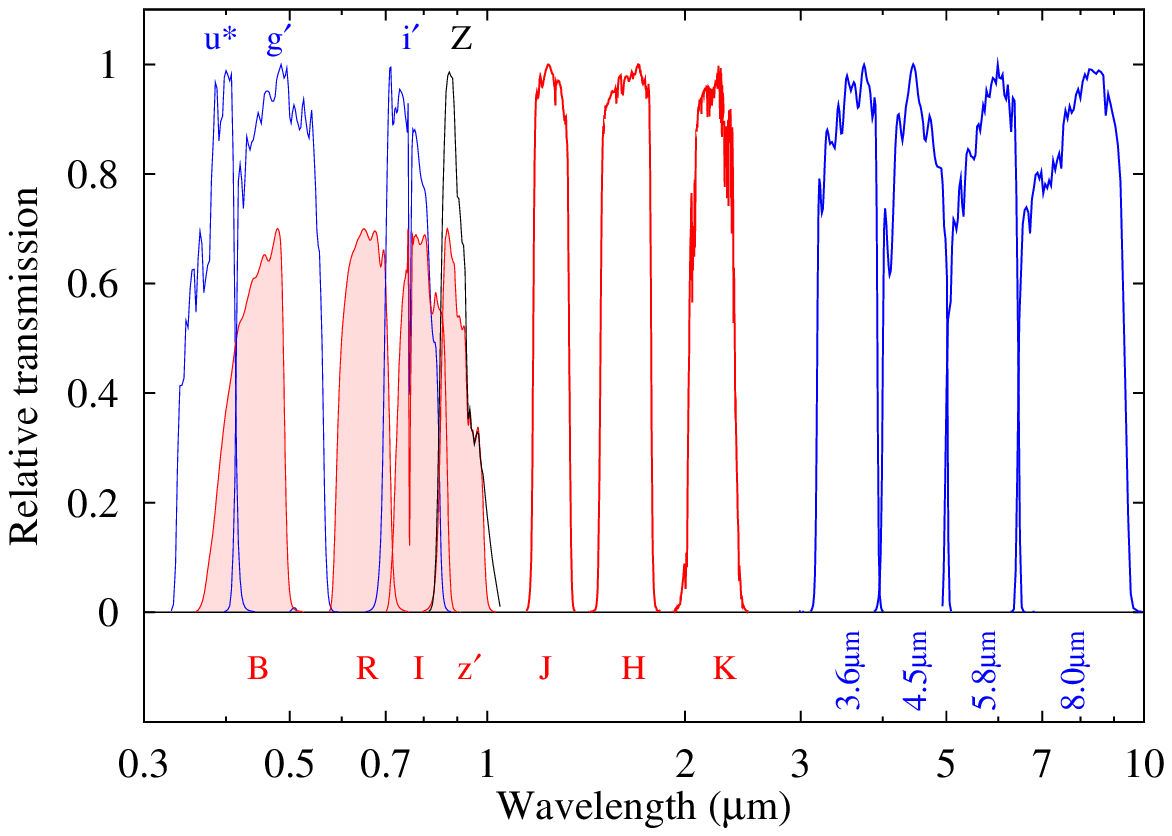}
\end{center}
\caption{The effective system transmission curves for each band available
for photometric redshifts in the 13$^{H}$ field. The transmission
curves have been normalised to have a peak transmission of unity,
except for the Subaru $BRIz'$ bands which, for clarity, are normalised to a peak
of 0.7 (and shaded).  }
\label{fig:passbands}
\end{figure}

For the IRAC data, the filter transmission curves for the 3.6, 4.5,
5.8 and 8.0 $\mu$m bands were downloaded from the Spitzer 
website\footnote{http://ssc.spitzer.caltech.edu/irac/rsrf/}
A plot of the transmission curves for the photometric filter set used in this paper is 
shown in figure \ref{fig:passbands}. It can be seen that the entire range 0.3--8
$\mu$m is well sampled with the largest gap lying between the 
NIR and IRAC coverage.

\subsection{Fine tuning of photometric zeropoints}
\label{sec:photo_calib}
A crucial input for accurate and reliable photometric redshifts is to
have accurately calibrated multi-band photometry. Even small relative
offsets in the zeropoints of individual bands can significantly
degrade the mean photometric accuracy and increase the rate of
catastrophic redshift errors (classed here as $|z_{photo} -
z_{spec}|/(1+z_{spec}) > 0.2$).  Therefore, to fine tune the relative
zeropoints of the optical/NIR imaging data we have devised a cross
calibration method using SDSS-DR6 and 2MASS photometric measurements of
stellar objects within the 13$^{H}$ field. This method exploits the
excellent photometric fidelity of the SDSS-DR6 and 2MASS surveys. In
particular we exploit the `Ubercalibrated' magnitudes from the SDSS-DR6,
which are absolutely calibrated to $<0.02$ mags \citep{ivezic07}.

Our procedure is to use the flux measurements of stars in the SDSS-DR6
$ugriz$ and 2MASS $JHK_S$ bands to predict the magnitudes of these
stars in the \filtsONIR\ science filter set (using a library of
template and synthetic stellar spectra). These reference magnitudes
are matched to the measured fluxes of these stars measured from the
science images, and then it is then a simple task to find the best
fitting zeropoint solution in each filter.  This method guarantees
that the science images themselves have been calibrated to a standard
photometric system rather than deriving the calibration from e.g. a
set of observations of standard stars that may have been observed at
a range of airmasses and photometric conditions.

Firstly we select all stellar objects from the SDSS-DR6 catalogue that
are covered by our science imaging of the 13$^{H}$ field.  We apply the
corrections (as suggested on the DR6 release
notes\footnote{http://www.sdss.org/dr6/start/aboutdr6.html}) to the
SDSS magnitudes to bring them into the AB system: $u'_{AB} = u'_{SDSS}
-0.04$ and $z'_{AB} = z'_{SDSS} +0.02$. We use the `psfMags' that are
appropriate for stellar objects.  We select the subset of stars that
have no SDSS flags set, that have $r' < 22.5$ and that have at least
one filter measurement with an error less than 0.05 mags. We calculate
the colours of each SDSS star relative to its `reference' magnitude
(defined as the mean of the magnitudes in the $g'$, $r'$ and $i'$
filters).
  We then choose a
standard stellar spectral template that best matches the measured
colours of each SDSS star, by folding the stellar template through the
SDSS transmission curves. We initially adopted a set of stellar
spectra from the Pickles (1998) and 
Bruzual-Persson-Gunn-Stryker (BPGS)\footnote{https://www.stsci.edu/hst/instrumentation/reference-data-for-calibration-and-tools/astronomical-catalogs/bruzual-persson-gunn-stryker-atlas-list} stellar spectral
libraries. However, we found that the locus in colour-colour space
occupied by the SDSS stars in the 13$^{H}$ field was systematically
offset (i.e. bluer in $u'-g'$ for a given $g'-r'$) from the colours
of the BPGS/Pickles reference stars.  This effect has been noted
before in SDSS studies \citep[e.g.][]{ivezic07}, where the measured
location of the stellar locus in colour-colour space moves at a rate
of $\sim$0.01 mag mag$^{-1}$. In high Galactic latitude fields, at
the faintest magnitudes ($r' \sim 22$) the reference stars are
presumably located in the Galactic halo, and hence have lower than
solar metallicity.  To allow for this we supplemented our stellar
template library with synthetic stellar templates from \citet{castelli03}, 
which span a wide range of
metallicities, and which adequately span the colour-colour locus of
the faint stars.  
The best fitting stellar template is folded through the \filtsONIR\ 
filter bandpasses (see section \ref{sec:filter_bandpasses}) to give
the expected magnitudes of each SDSS reference star in each band.

Zeropoints in each filter are then
determined by a weighted least squares fit to the measured image
fluxes. This process results not only in a self consistent calibration
of magnitudes across all filters, but also good estimates of the
residual calibration uncertainties.

We note that this process is similar to the 
method used to calibrate images in the MegaPipe pipeline, except that we 
calculate colour equations from stellar templates
on a star by star basis rather than using empirical colour
equations derived for a nominal stellar population (that may not be 
representative of the observed stars).

\label{lastpage}

\end{document}